\documentclass[12pt,sort&compress]{elsarticle}
\usepackage[utf8]{inputenc}

\usepackage{bm}
\usepackage{placeins}

\usepackage{mathrsfs}
\usepackage{amsmath}
\usepackage{amsfonts}
\usepackage{amsthm}
\usepackage{upgreek}
\usepackage{color}
\usepackage{sansmath}

\usepackage{array}

\usepackage{floatrow} 
\usepackage{graphicx}
\usepackage[tableposition=top]{caption}

\usepackage{caption}
\usepackage{subcaption}
\usepackage{float}
\usepackage{url}
\usepackage{multicol}
\usepackage[top=1in,bottom=1in,left=1in,right=1in]{geometry}
\usepackage[pdfborder={0 0 0},colorlinks,allcolors=blue]{hyperref}
\usepackage{txfonts}
\usepackage{setspace}
\usepackage{arydshln}
\usepackage{enumitem}
\usepackage{lipsum}
\usepackage{siunitx,booktabs}
\usepackage{nth}
\usepackage{accents} 

\usepackage{float}  
\usepackage{floatrow} 

\theoremstyle{definition}

\allowdisplaybreaks

\usepackage{prettyref}
\newrefformat{fig}{Figure~\ref{#1}}
\newrefformat{Fig}{Fig.~\ref{#1}}
\newrefformat{Eq}{Eq.(\ref{#1})}
\newrefformat{eq}{Eq.(\ref{#1})}
\newrefformat{sec}{Section~\ref{#1}}
\newrefformat{tab}{Table~\ref{#1}}
\newrefformat{Tab}{Table~\ref{#1}}
\newrefformat{alg}{Algorithm~\ref{#1}}
\newrefformat{Alg}{Algorithm~\ref{#1}}

\usepackage[dvipsnames]{xcolor}
\newcommand\hl[1]{%
	\bgroup
	\hskip0pt\color{red!80!black}%
	#1%
	\egroup
}

\usepackage{algorithm}
\usepackage{algpseudocode}
\usepackage{algorithmicx}
\definecolor{shadecolor}{cmyk}{0,0,0,0.03}

\usepackage{amssymb}

 \usepackage{lineno}
\journal{Physics of Fluids}

\begin{document}
\begin{frontmatter}




\title{{\normalsize\textit{This article has been submitted to Physics of Fluids.} \url{https://aip.scitation.org/journal/PoF}}\\[0.5em]
A novel machine learning framework based on proper orthogonal decomposition for turbine wake flow predictions in natural waterways}

\author[sbu]{Mustafa Meriç Aksen}
\author[sbu]{Hossein Seyedzadeh}
\author[sbu]{Mehrshad Gholami Anjiraki}
\author[sbu]{Ali Khosronejad\corref{cor1}}
\ead{ali.khosronejad@stonybrook.edu}
\cortext[cor1]{Corresponding author}

\address[sbu]{Department of Civil Engineering, Stony Brook University, Stony Brook, NY 11794, USA}

\begin{abstract}
\noindent We developed a machine-learning (ML) surrogate trained on instantaneous large-eddy simulation (LES) data filtered using proper orthogonal decomposition (POD) to predict mean flow fields in natural rivers with complex geometries. The surrogate enables efficient prediction of mean flow fields downstream of marine hydrokinetic (MHK) turbine arrays at a fraction of the computational cost of high-fidelity simulations, facilitating turbine-layout optimization in natural waterways where bathymetry, turbulence, and turbine interactions influence the available energy resource. We demonstrate the model’s capabilities at a real-world tidal river site under ebb-tide conditions. Complex bed bathymetry and planform geometry generate substantial background turbulence that can obscure turbine-wake signatures. POD is employed to extract relevant flow structures, improving the ML model’s ability to learn wake dynamics. First, LES is performed at the tidal site to generate training data. POD is then applied to decompose the instantaneous LES fields into spatial modes. Selected modes are used to train a three-dimensional convolutional encoder–decoder network to predict time-averaged velocity fields. This selective filtering reduces the influence of dominant background fluctuations while preserving information relevant to turbine wakes. At equivalent network size and training cost, the POD-filtered surrogate reduces the mean-squared error by three orders of magnitude compared with the same network trained on unfiltered LES data.
\end{abstract}


\end{frontmatter}

\section{Introduction}

\label{sec:intro}

\noindent Tidal energy requires site-specific resource assessments that evaluate the available resource alongside the technical, environmental, and economic constraints on energy extraction \cite{[2]}. High-fidelity computational fluid dynamics (CFD), particularly large-eddy simulation (LES), provides detailed representations of turbulent flow and turbine–wake interactions that support tidal-energy assessment and array design \cite{[8],[9]}. \citet{AKSEN2026105302} studied the tidal flows of Piscataqua River, New Hampshire, and established the feasibility of tidal energy extraction at this site, while identifying hypothetical candidate locations with high energy potential. First, before installing (virtually) turbine arrays, they employed site-scale LES to characterize the ambient hydrodynamics of a 6km-long reach of the tidal river under ebb and flood conditions, requiring on average approximately 20,000 processor-hours per simulation to reach statistically converged flow fields \cite{AKSEN2026105302}. Extending this framework to systematically optimize turbine placements would require repeated LES of turbine-induced wakes and array interactions with numerous candidate configurations, which makes the optimization of turbine array layouts computationally prohibitive \cite{[7],[10]}.

Machine learning (ML) models offer a broad framework for extracting information from numerical simulation data, supporting flow field modeling, and optimizing turbine array layouts \cite{[11]}. In particular, selected CFD results can be used to construct computationally inexpensive approximations of array performance \cite{[10]}. For example, \citet{w18121500} implemented an ML model to predict the mean flow field of the Piscataqua River using a limited set of instantaneous LES data during the training process. The idea behind this method is to use a limited set of instantaneous LES data to train the ML model. The trained ML model is then used to predict the time-averaged (mean) flow field with a computational cost less than $1\%$ of that of LES \cite{CNN_1, CNN_2, CNN_3, CNN_4}. 
Building on these past works, the present study aims to train an ML model such that it can identify the signature of turbine wake dynamics in natural rivers with overwhelming background turbulent flow structures. The trained ML model is then employed to optimize tidal turbine layouts, maximizing annual energy production at a tidal site, where the turbulent flow is governed primarily by bathymetry- or planform-induced turbulence rather than turbine-induced turbulence. Although neural networks, including physics-informed variants, can learn and predict complex flow behavior, the highly complex turbulent flow of natural rivers challenges their ability to represent small-scale eddies accurately, thereby affecting turbine-induced structures.

Reduced-order models address this challenge differently. Rather than approximating every scale of the flow, they seek to describe the flow with a specific frequency, scale, or energy. Their objective is to approximate a high-dimensional system using a limited set of coordinates while retaining the behavior relevant to analysis, prediction, or optimization \cite{Benner2015}. These models are employed in control systems \cite{Moore1981}, structural dynamics \cite{CRAIG1968}, and fluid mechanics, which are also related to the present study, through reduced-order models that describe coherent flow structures and support feedback control \cite{Rowley2017}. Among modal approaches to reduced-order modeling, proper orthogonal decomposition (POD) is widely used to extract coherent structures and construct compact representations of turbulent flows, with the snapshot method making its application practical for high-dimensional datasets \cite{Sirovich1987,Berkooz1993}. In theory, POD constructs an orthogonal basis that minimizes the mean-square reconstruction error of the sampled fields under a specified inner product, equivalently capturing the maximum energy for a given number of modes. However, minimizing global reconstruction error does not necessarily preserve the flow features that are most relevant to turbine-turbulence interactions. Evidence for the importance of the prediction objective comes from wind turbine wake analysis. 

More specifically, \citet{Bastine2015} demonstrated that the relevance of individual POD modes depends on the downstream quantity being reconstructed, and that selected wake observables can be represented with relatively few modes despite recovering only a limited fraction of the turbulence kinetic energy. Higher-order modes should therefore be assessed according to their physical contribution rather than their energetic rank alone. Moreover, the energetic ranking of spatial POD modes does not impose a corresponding frequency ordering, because individual spatial POD modes can contain contributions from a wide range of frequencies \citep{Towne2018}. Several POD variants supplement conventional energy-based representations: balancing POD to incorporate controllability and observability for linear input--output systems \citep{ROWLEY2005}. Spectral POD identifies structures that evolve coherently at individual frequencies in statistically stationary flows \citep{Towne2018}, and multiscale POD separates contributions across prescribed temporal-frequency scales \citep{Mendez2019}. The distinction between energetic importance and predictive relevance matters for marine energy turbine arrays, where experiments and LES results show that ambient turbulence, wake asymmetry, and turbine arrangement influence wake interactions and downstream performance \citep{Churchfield1985,MYCEK2014876}. 

Although these variants change how modes are selected and ranked, they still represent the flow as a linear superposition of modes. This reliance on a linear subspace limits the ability of POD-based models to represent strongly nonlinear dynamics. In highly turbulent flows such as riverine flows that are strongly influenced by site-specific characteristics, wake structures relevant to turbine interactions may require a significant number of modes. One remedy (to the excessive number of required modes) is to enrich the basis with additional physically informed modes. For example, incorporating a shift mode substantially improved predictions of transient vortex-shedding dynamics in the cylinder wake \citep{Noack2003}. In complex turbulent flows such as those encountered in this work, however, the physically important dynamics may not be identifiable in advance, and no fixed set of linearly superposed modes may represent them efficiently. Nonlinear manifold approaches based on convolutional encoder-decoder networks have therefore emerged as a promising alternative.

Convolutional autoencoders could construct reduced-order representations of high-dimensional turbulent flow fields by learning nonlinear mappings between the original flow variables and a compact latent space, while their encoder–decoder structure enables spatial compression and reconstruction, supporting both flow-feature extraction and reduced-order modeling \cite{Murata2020,LEE2020108973}. For turbulent inflow generation, autoencoder-type convolutional networks have been used to generate time-dependent inlet velocity fields that sustain turbulent channel flow at substantially lower computational cost than the corresponding driver simulations \cite{Fukami2019}. Extensions to three-dimensional (3D) turbulence have combined convolutional compression with recurrent networks to represent spatial structures and their temporal evolution. \citet{Mohan02102020} developed a compressed convolutional long short-term memory model for the 3D homogeneous isotropic turbulence and evaluated its performance using physics-based statistical diagnostics, identifying both capabilities and limitations. Nonetheless, the need for a relatively large number of latent modes was identified as a limitation for extending the approach to more complex flows, such as higher-Reynolds-number turbulence \cite{Nakamura2021}. The generalization to previously unseen conditions remains sensitive to the extent of the coverage of the training data. In that regard, the ML model is expected to perform better when test conditions remain close to those represented during training \cite{Morimoto2022}. Retention of physical information during the compression phase is important \cite{Fukami2023}.

In the context of marine hydrokinetic (MHK) energy applications, we hypothesize that energetic background fluctuations induced by complex river bathymetry and/or its planform geometry may obscure the comparatively localized flow signatures associated with turbine wakes during the learning process. In particular, the large-scale and spatially repeating flow structures generated by bathymetric variations may dominate the flow-field variability, making it more difficult for a neural network to identify and learn the characteristic features of turbine-induced wake dynamics. Motivated by this hypothesis, we investigate whether bathymetry-related large-scale, recurring flow structures can be systematically filtered or truncated from the neural-network input to focus the learning process more directly on the flow features associated with the turbine wake. We then assess the extent to which this targeted representation affects the model's ability to reconstruct and predict the turbine wake. Specifically, we evaluate the model's performance both qualitatively, through its ability to reproduce the spatial structure and evolution of the wake, and quantitatively, using appropriate flow-field error metrics. This approach allows us to determine whether isolating wake-relevant flow structures from the broader bathymetry-induced variability can improve the accuracy and robustness of ML-based wake prediction in complex natural environments.

Building on this hypothesis, herein, we first generate and curate a high-fidelity LES dataset comprising multiple tidal-turbine layout scenarios at the candidate site proposed by \citet{AKSEN2026105302} under ebb-tide conditions. We then investigate whether POD applied to instantaneous velocity fields can effectively separate and filter bathymetry-associated flow structures, thereby providing a physics-informed preprocessing step that enables the neural network to focus more directly on turbine-wake dynamics. Using the resulting flow-field representation, we develop a convolutional encoder–decoder surrogate model and evaluate its ability to provide stable and accurate predictions in the presence of strongly heterogeneous, 3D hydrodynamics and multiple installed turbines. Particular emphasis is placed on determining whether removal of dominant bathymetry-related structures improves the model's ability to resolve the comparatively localized and dynamically evolving signatures of turbine wakes.

The Piscataqua River is selected as the case-study site because Portsmouth Harbor was identified by the National Laboratory of the Rockies as a tidal-energy hotspot, ranking ninth in short-term and eleventh in long-term investment potential \cite{NREL}. Beyond evaluating a novel ML framework for tidal-wake prediction, the present study provides a detailed assessment of POD-based flow decomposition in a realistic riverine environment characterized by complex bathymetry and interacting turbine wakes. The resulting analysis offers insight into the relative contributions of bathymetry- and turbine-induced flow structures and their implications for data-driven modeling. More broadly, the findings provide transferable guidance for the application of LES, reduced-order modeling, and ML-based surrogate models to tidal-turbine array design and performance assessment in constrained and hydrodynamically heterogeneous natural waterways.

The subsequent sections of the paper are organized as follows. \prettyref{sec:mathematical_background} presents the mathematical background for the hydrodynamic governing equations, POD, and describes the POD-filtered 3D U-Net (POD-U-Net) framework as the workhorse ML model of this study. Next, we describe in \prettyref{sec:pod} the results of the POD. In \prettyref{sec:rad}, we compare the results of the POD-U-Net with the high-fidelity LES data. Lastly, \prettyref{sec:conc} summarizes the significance of the findings and their implications for future research.

\section{Mathematical Background}
\label{sec:mathematical_background}

\noindent Herein, we summarize the LES formulation, bathymetry treatment, and turbine parametrization method used to generate the wake flow field data. Repeated indices imply summation; $u_l$ and $p^\star$ denote spatially filtered Cartesian velocity and modified pressure. An overbar denotes spatial filtering, whereas $\langle\cdot\rangle_t$ denotes temporal averaging in Sections~\ref{sec:pod} and~\ref{sec:pod_unet}.

\subsection{Hydrodynamic's governing equations}
\label{sec:hydrodynamic_equations}
\noindent The incompressible Navier--Stokes equations are solved in generalized curvilinear coordinates \citep{Ge2007Curvilinear,kang2011high}. For a stationary mapping, the Jacobian, transformation metrics, and contravariant volume flux are
\begin{equation}
J=\det\left(\frac{\partial\boldsymbol\xi}{\partial\mathbf x}\right),
\qquad \xi_l^i=\frac{\partial\xi^i}{\partial x_l},
\qquad G^{jk}=\xi_l^j\xi_l^k,
\qquad U^i=\frac{\xi_l^i}{J}u_l.
\label{eq:coordinate_metrics}
\end{equation}
where $\xi_l^i=\partial\xi^i/\partial x_l$, $J$ maps physical to computational coordinates, and $G^{jk}$ is the contravariant metric tensor. For a constant-density Newtonian fluid, conservation of mass and momentum take the form:
\begin{equation}
J\frac{\partial U^j}{\partial\xi^j}=0,
\label{eq:les_continuity}
\end{equation}
\begin{equation}
\begin{aligned}
\frac{\partial U^i}{\partial t}
=\xi_l^i\Bigg[&
-\frac{\partial(U^ju_l)}{\partial\xi^j}
-\frac{1}{\rho}\frac{\partial}{\partial\xi^j}
 \left(\frac{\xi_l^jp^\star}{J}\right)
+\frac{1}{\rho}\frac{\partial}{\partial\xi^j}
 \left(\mu\frac{G^{jk}}{J}
 \frac{\partial u_l}{\partial\xi^k}\right)\\
&-\frac{1}{\rho}\frac{\partial}{\partial\xi^j}
 \left(\frac{\xi_m^j}{J}\tau_{lm}\right)
+\frac{f_l}{J}\Bigg].
\end{aligned}
\label{eq:les_momentum}
\end{equation}
where $\rho$ is the density of fluid, $\mu$ is dynamic viscosity of fluid, $p^\star=p+\tau_{kk}/3$ includes the isotropic subgrid-scale (SGS) stress, $p$ excludes the hydrostatic contribution, and $f_l$ is Cartesian body force per unit mass. The SGS stress is
\begin{equation}
\tau_{lm}=
\rho\left(\overline{v_lv_m}-u_lu_m\right),\qquad u_l=\overline{v_l}.
\label{eq:sgs_definition}
\end{equation}
where $v_l$ is the unfiltered velocity, and the overbar denotes spatial filtering, and the deviatoric SGS stress is modeled using the dynamic Smagorinsky closure \citep{Smagorinsky1963,Germano1991Dynamic}:
\begin{equation}
\tau_{lm}-\frac{1}{3}\tau_{kk}\delta_{lm}
=-2\mu_t\overline S_{lm},
\qquad \mu_t=\rho C_{\mathrm{s}}\Delta^2|\overline{\mathbf S}|,
\label{eq:sgs_closure}
\end{equation}
where $\delta_{lm}$ is the Kronecker delta. The Cartesian strain-rate components are evaluated using curvilinear derivatives:
\begin{equation}
\overline S_{lm}=\frac{1}{2}\left(
\xi_m^j\frac{\partial u_l}{\partial\xi^j}
+\xi_l^j\frac{\partial u_m}{\partial\xi^j}\right),
\qquad |\overline{\mathbf S}|=(2\overline S_{lm}\overline S_{lm})^{1/2}.
\label{eq:resolved_strain}
\end{equation}
The dynamic procedure determines $C_{\mathrm{s}}$ from the resolved field \citep{Germano1991Dynamic}; $\mu_t$ is the dynamic eddy viscosity. Following the cell-volume filter used in the flow solver \citep{kang2011high}, the filter width is
\begin{equation}
\Delta=(\Delta V)^{1/3},
\qquad \Delta V=\int_{\mathrm{cell}}J^{-1}\,\mathrm d^3\xi.
\label{eq:filter_width}
\end{equation}
where the unit computational cell increment $\Delta\simeq J^{-1/3}$, and for more details concerning the mathematical background of turbulence modeling in this study, the reader is referred to \cite{GholamiAnjiraki2025Vertical, Seyedzadeh2023Lagrangian, AKSEN2025122110}.

\subsection{Proper Orthogonal Decomposition
}
\label{sec:pod}
\noindent POD extracts the most energetic spatial patterns from a time-resolved flow field and represents the flow using a reduced set of orthogonal modes \citep{Sirovich1987Snapshots,Taira2017Overview}. For the present velocity field, the instantaneous fluctuation can be expressed as:
\begin{equation}
  \mathbf{u}'(\mathbf{x},t)
  =\mathbf{u}(\mathbf{x},t)-\langle\mathbf{u}\rangle_t(\mathbf{x})
  \approx \sum_{i=1}^{N_r} a_i(t)\,\boldsymbol{\phi}_i(\mathbf{x}),
  \label{eq:vel-fluc}
\end{equation}
where $\langle\mathbf{u}\rangle_t$ is the supplied time-averaged velocity, $\boldsymbol{\phi}_i$ is the $i$th spatial mode, $a_i$ is its temporal coefficient, and $N_r$ is the number of retained modes. We note that since the temporal mean is subtracted before performing POD, mode 1 is the leading fluctuation mode rather than the mean field.

\subsubsection{Method of snapshots}
\noindent The method of snapshots \citep{Sirovich1987Snapshots} considers $m$ flow realizations sampled at discrete times, with $m\ll n$, where $n$ is the number of spatial degrees of freedom. It therefore replaces the classical $n\times n$ spatial eigenvalue problem with an $m\times m$ temporal eigenvalue problem. The snapshots must cover a sufficiently long period and be sampled frequently enough to resolve the energetic fluctuations of interest \citep{Taira2017Overview}.

Herein, the Cartesian velocity components $u$, $v$, and $w$ are treated jointly. Values stored at grid nodes are first interpolated to cell centers by averaging the eight vertices surrounding each cell. The separately calculated temporal mean is then subtracted from every snapshot. After flattening the three cell-centered components, snapshot $j$ is represented by
\begin{equation}
  \mathbf{q}_j=
  \begin{bmatrix}
  u'_1 & \cdots & u'_{N_c} &
  v'_1 & \cdots & v'_{N_c} &
  w'_1 & \cdots & w'_{N_c}
  \end{bmatrix}^{T},
  \label{eq:pod-snapshot-vector}
\end{equation}
where $N_c$ is the number of cell-centered grid points included in the analysis. The snapshot matrix is
\begin{equation}
  \mathbf{X}=\begin{bmatrix}\mathbf{q}_1&\mathbf{q}_2&\cdots&\mathbf{q}_m\end{bmatrix}
  \in\mathbb{R}^{3N_c\times m}.
  \label{eq:pod-snapshot-matrix}
\end{equation}

To account for the nonuniform curvilinear grid and exclude solid cells, the inner product is weighted by the normalized fluid-cell volumes, as:
\begin{equation}
  \langle\mathbf{a},\mathbf{b}\rangle_{\mathbf M}
  =\mathbf{a}^{T}\mathbf{M}\mathbf{b},
  \label{eq:pod-weighted-product}
\end{equation}
where $\mathbf{M}$ is a diagonal spatial-weight matrix that applies the same normalized cell-volume weight to the three velocity components. This volume-weighted inner product is the discrete counterpart of the kinetic-energy inner product used in POD \citep{Taira2017Overview}. The temporal correlation matrix is then formed as:
\begin{equation}
  \mathbf{C}=\frac{1}{m}\mathbf{X}^{T}\mathbf{M}\mathbf{X},
  \qquad
  \mathbf{C}\boldsymbol{\psi}_i=\lambda_i\boldsymbol{\psi}_i.
  \label{eq:pod-eigenproblem}
\end{equation}
This formulation requires the solution of an $m\times m$ eigenvalue problem rather than the much larger spatial problem. The eigenvalues are ordered as $\lambda_1\geq\lambda_2\geq\cdots\geq0$, so that the modes are ranked from the greatest to the least energetic contribution.

The spatial modes of the POD and their temporal coefficients are recovered from the temporal eigenvectors as
\begin{equation}
  \boldsymbol{\phi}_i
  =\frac{1}{\sqrt{m\lambda_i}}\mathbf{X}\boldsymbol{\psi}_i,
  \qquad
  a_i(t_j)=\sqrt{m\lambda_i}\,\psi_{ji},
  \label{eq:pod-modes-coefficients}
\end{equation}
and satisfy $\boldsymbol{\phi}_i^{T}\mathbf{M}\boldsymbol{\phi}_j=\delta_{ij}$. This construction is equivalent to applying singular-value decomposition to the weighted snapshot matrix $\mathbf{M}^{1/2}\mathbf{X}$ \citep{Taira2017Overview}. The relative and cumulative modal contributions are
\begin{equation}
  E_i=\frac{\lambda_i}{\sum_{j=1}^{m}\lambda_j},
  \qquad
  E_{1:r}=\frac{\sum_{i=1}^{r}\lambda_i}{\sum_{j=1}^{m}\lambda_j}.
  \label{eq:pod-energy}
\end{equation}
The code can retain either a prescribed number of modes or the smallest number required to reach a specified cumulative contribution. In the present analysis, modes 1--45 were considered. For the POD-filtered network introduced in Section~\ref{sec:pod_unet}, the contribution of a selected modal set $\mathcal{K}$ can be removed (filtered) from each instantaneous field, as follows:
\begin{equation}
  \mathbf{u}_{\mathrm{F}}(\mathbf{x},t)
  =\mathbf{u}(\mathbf{x},t)
  -\sum_{i\in\mathcal{K}}a_i(t)\boldsymbol{\phi}_i(\mathbf{x}).
  \label{eq:pod_selective_filter}
\end{equation}
As discussed later in this paper, herein, the modal sets that are filtered out are consist of $\mathcal{K}=\{2,\ldots,7\}$. This selective subtraction differs from a conventional low-rank reconstruction in that  only the mean, mode 1, modes 8--45, and any contribution outside the retained POD subspace remain in the considered flow field data.

\subsection{POD-filtered convolutional neural network scheme}
\label{sec:pod_unet}

\noindent The proposed framework, hereafter POD-U-Net, combines the selective POD filter in Eq.~\eqref{eq:pod_selective_filter} with a 3D convolutional encoder--decoder. Convolutional neural networks (CNNs) have been used to learn spatial mappings of turbulent flow fields at substantially lower evaluation cost than high-fidelity simulation \citep{Brunton2020MLFluids,ZexiaPECNN,w18121500}. Here, the network maps an instantaneous POD-filtered velocity field to the corresponding time-averaged velocity field before performing selective modal subtraction.

\begin{figure}[htbp]
  \centering
  \includegraphics[width=\textwidth]{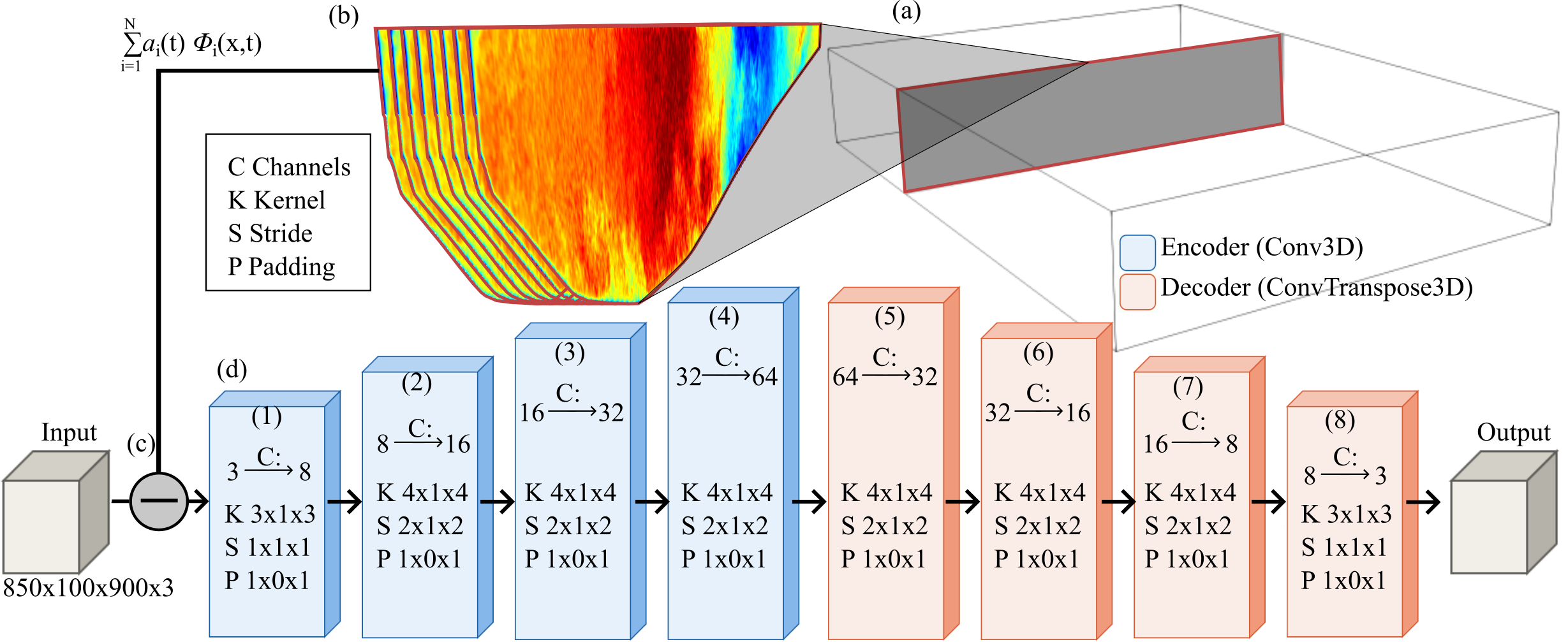}
  \caption{POD-U-Net framework for predicting the 3D time-averaged velocity field. (a) Extraction of the 3D LES velocity field from the domain; (b) POD representation of the dominant flow structures; (c) subtraction of the selected modal contribution, corresponding to modes 2--7 in the present analysis; and (d) convolutional encoder--decoder that maps the filtered instantaneous input to the predicted time-averaged velocity field. $C$, $K$, $S$, and $P$ denote the number of channels, kernel size, stride, and padding, respectively.}
  \label{Fig:pod_cnn_framework}
  \label{fig:pod_cnn_framework}
\end{figure}

Figure~\ref{fig:pod_cnn_framework} describes the complete procedure. Each instantaneous LES-computed velocity field contains the spatially filtered Cartesian components $\mathbf{u}=(u,v,w)^T$. After interpolation to the cell-centered grid, the modal contribution associated with $\mathcal K=\{2,\ldots,7\}$ is removed. The resulting field $\mathbf{u}_{\mathrm F}$ forms the three-channel network input. The prediction target is the temporal mean $\langle\mathbf{u}\rangle_t$ before selective modal subtraction; therefore, the removed modes are not added back to the network output.

\subsubsection{Input--output mapping}

\noindent For layout $\ell$, the LES target is
\begin{equation}
  \left\langle\mathbf{u}^{(\ell)}\right\rangle_t
  =\frac{1}{N_{\mathrm{avg}}^{(\ell)}}
  \sum_{s=1}^{N_{\mathrm{avg}}^{(\ell)}}
  \mathbf{u}^{(\ell,s)},
  \label{eq:cnn_target}
\end{equation}
where $N_{\mathrm{avg}}^{(\ell)}$ is the number of LES fields used for temporal averaging. Let $\mathcal S_{\mathrm{tr}}$ denote the scaling operator fitted using the training set and $\mathcal G_{\boldsymbol\theta}$ the CNN with parameters $\boldsymbol\theta$. The predicted temporal mean is
\begin{equation}
  \widehat{\left\langle\mathbf{u}^{(\ell)}\right\rangle_t}
  =\mathcal S_{\mathrm{tr}}^{-1}\!\left[
  \mathcal G_{\boldsymbol\theta}\!\left(
  \mathcal S_{\mathrm{tr}}\!\left[\mathbf{u}_{\mathrm F}^{(\ell,s)}\right]
  \right)\right].
  \label{eq:cnn_mapping}
\end{equation}
The same training-derived normalization is applied to all validation and test inputs, following standard supervised-learning practice \citep{Goodfellow2016DeepLearning}. The mean field, POD basis, modal set, and normalization must remain fixed when the trained model is evaluated on a held-out layout.

\subsubsection{Network architecture and training procedure}

\noindent The architecture shown in Fig.~\ref{fig:pod_cnn_framework} follows a volumetric convolutional encoder--decoder design related to 3D U-Net architectures \citep{Cicek2016UNet}. The encoder contains four convolutional stages with channel progression $3\rightarrow8\rightarrow16\rightarrow32\rightarrow64$, followed by four transposed-convolution stages with the reverse progression. The first and final stages use $3\times1\times3$ kernels, unit stride, and $1\times0\times1$ padding. The intermediate stages use $4\times1\times4$ kernels, $2\times1\times2$ stride, and $1\times0\times1$ padding. Thus, the network downsamples and upsamples in the first and third grid directions while retaining the resolution in the second direction. The architecture shown in Fig.~\ref{fig:pod_cnn_framework} contains no skip connections.

The network is trained by minimizing the mean-squared error over the fluid cells and three velocity components \citep{Goodfellow2016DeepLearning}:
\begin{equation}
  \mathcal L(\boldsymbol\theta)
  =\frac{1}{3N_{\mathrm{tr}}N_c}
  \sum_{s=1}^{N_{\mathrm{tr}}}
  \sum_{c=1}^{N_c}
  \sum_{\alpha=1}^{3}
  \left[
  \widehat{\left\langle u_{\alpha,c}^{(s)}\right\rangle_t}
  -\left\langle u_{\alpha,c}^{(s)}\right\rangle_t
  \right]^2,
  \label{eq:cnn_loss}
\end{equation}
where $N_{\mathrm{tr}}$ is the number of training samples and $N_c$ is the number of fluid cells. An unfiltered CNN with the same architecture, targets, loss, and training settings is used as the baseline. Consequently, differences between the two models isolate the effect of POD filtering; the filtering operation does not alter the number of trainable network parameters.

The predictions are converted to physical units before comparison with the LES temporal mean. Performance is evaluated using the training and testing loss histories, spatial velocity maps, cross-sectional contours, and velocity profiles through the turbine-array region. These comparisons quantify both domain-wide agreement and the recovery of turbine-induced near- and far-wake velocity deficits.

\subsection{Case study description and computational details}

\begin{figure} [H]
  \includegraphics[width=1\textwidth]{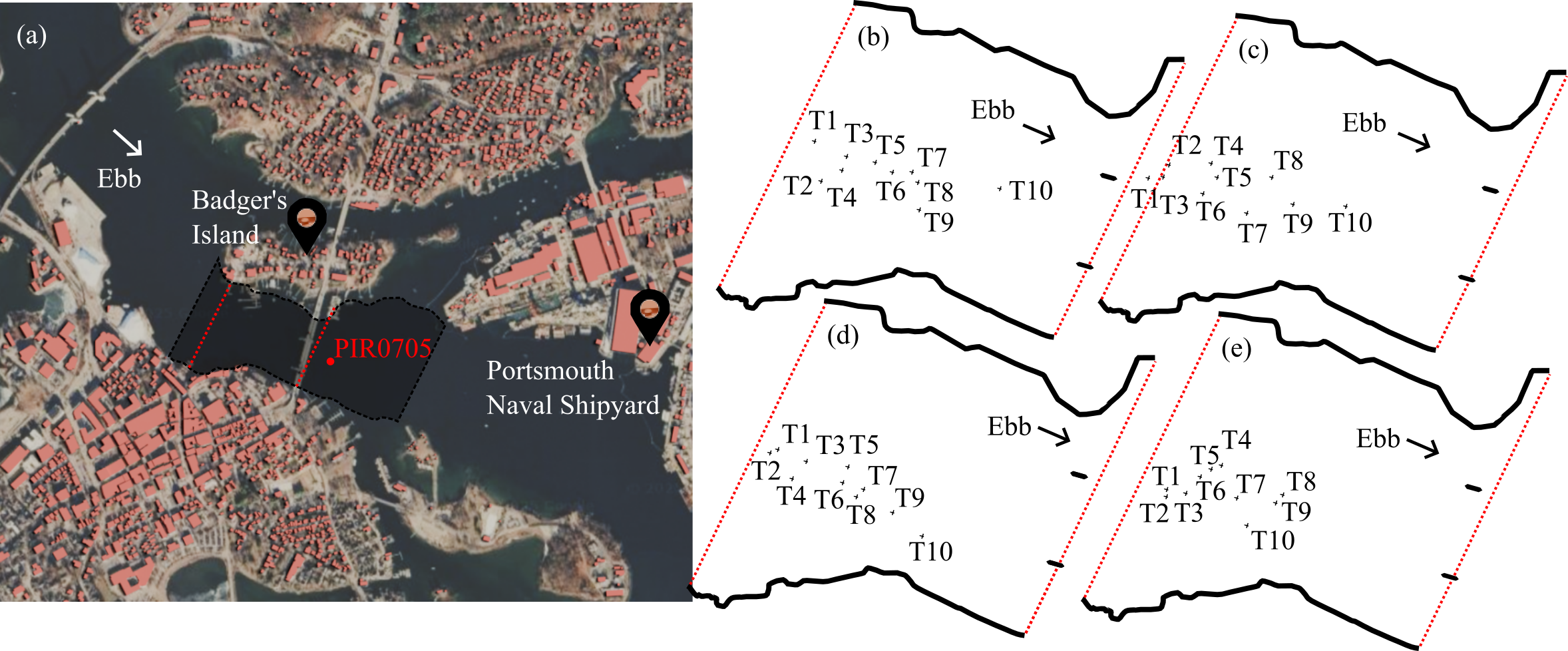} 
  \caption{Plan view of the computational domain as indicated by the gray region (a) and from (b) to (e) tidal layout scenarios 1, 2, 3, and 4 are demonstrated, respectively. The dashed gray line indicates the computational domain for LES, while the red dotted line shows the domain used for the machine learning and POD analysis.}
  \label{Fig:2}
  \label{fig:2}
\end{figure}

\noindent Following a series of past studies (see, e.g., \cite{AKSEN2026105302} and \cite{NREL}), the focus of this study is on a reach of the Piscataqua River in Portsmouth Harbor in the northeastern continental United States, which carries some of the fastest tidal currents on the U.S. East Coast \cite{USACE2014}. Importantly, the United States National Laboratory of the Rockies has ranked this study site to be among the top tidal energy sites for both short- and long-term investment in the country \cite{NREL}. A site-scale LES of this reach was validated against acoustic Doppler current profiler (ADCP) measurements near the Memorial Bridge, which is located within the study area \cite{AKSEN2026105302}. They reported that the river in the study area is strongly ebb-dominated \cite{AKSEN2026105302}. It also identified the reach between Badger's Island and the Portsmouth Naval Shipyard as the most favorable deployment site, as this reach can accommodate a 5 m rotor within the navigation constraints of the harbor's deep-draft channel (see \prettyref{Fig:2}). As seen, the flow domain, which is shown in blue in \prettyref{Fig:2}a, extends approximately $1~km$ in the streamwise direction and includes the Memorial Bridge piers and National Oceanic and Atmospheric Administration (NOAA) current station PIR0705. 

Following an array-scale grid-refinement study, which is reported in \citet{AKSEN2026105302}, a computational grid spacing of $D/20=0.25~m$ is utilized in all directions near the turbine blades. Away from the turbine array, the computational grid system is gradually stretched to a maximum resolution of  about $1~m$ in the streamwise and spanwise directions, while the vertical spacing is held at $0.25~m$. The resulting grid contains approximately 110 million computational cells, and all dimensions of the flow domain are normalized with the maximum water depth, $\mathrm{H_{max}}=25~m$. 

 We performed LES of the flow field for the four tidal turbine layouts plotted in  \prettyref{Fig:2}(b) to (e), under the ebb-tide condition at the site with a reference inflow velocity of $U_{\infty}=2.16$~m\,s$^{-1}$. This corresponds to the maximum-ebb event measured on May 1, 2017. The corresponding Reynolds number is $Re_{{H_{max}}} = 5.4 \times  10^7$. The water surface is represented by a rigid lid. The river bed and bridge piers are resolved with the curvilinear immersed boundary (CURVIB) method, and near-bed flow is modeled with a wall model \cite{wallmodel}. A time step of $\Delta t=0.02~s$ is selected to keep the Courant–Friedrichs–Lewy (CFL) number below unity. Inflow conditions are obtained by one-way nesting from the validated site-scale LES \cite{AKSEN2026105302}, i.e., at each time step, the instantaneous velocity field is interpolated in space and time and imposed at the inlet of the flow domain. 

As outlined by the red dotted line in \prettyref{Fig:2}a, the present study area lies within the narrow passage that was previously reported as the most favorable region in terms of momentum availability \cite{AKSEN2026105302}. Four layouts, each with ten turbines, are considered (see \prettyref{Fig:2}b--e). The turbine model is based on the Gen4 Kinetic Hydropower System (KHPS) developed by Verdant Power  \cite{[rev_re_4]}. It has a rotor diameter of $D=5\ \mathrm{m}$ and a hub height of 5 m above the local bed. The blades are represented by the actuator line method at a constant tip-speed ratio of $\lambda = 2$, and the nacelle by the actuator surface method \cite{Aksen2024Debris}.


\section{Proper Orthogonality Decomposition}

\label{sec:pod}

\subsection{Hydrodynamic Analysis}

\begin{figure} [H]
  \includegraphics[width=1\textwidth]{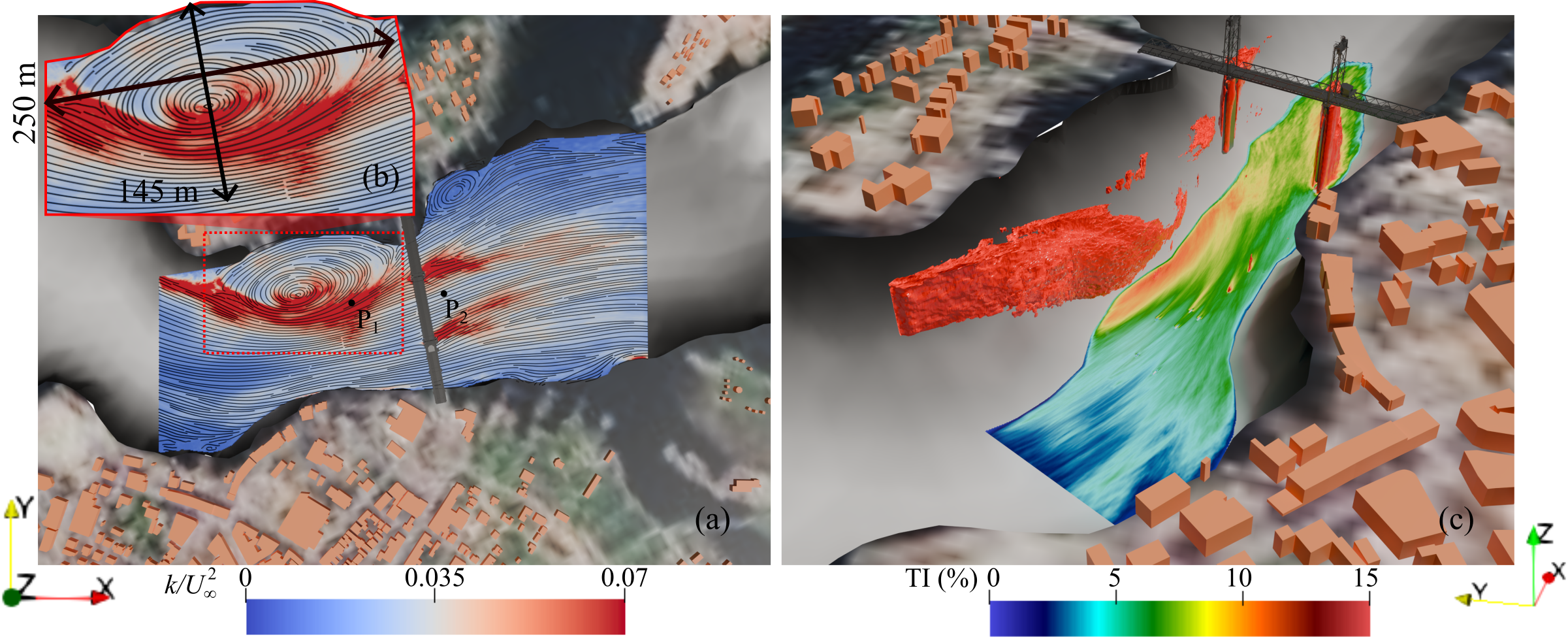} 
  \caption{LES-computed flow field of layout 3, with (a) plotting the streamlines of the time-averaged velocity field colored with contours of $k/U_{\infty}^2$ at the water surface. Points 'P1' and 'P2' mark the probe locations used for the spectral analysis in \prettyref{Fig:4}. (b) depicts the strong recirculation zone around Badger's Island. Additionally, (c) shows the color map of LES-computed turbulence intensity, TI (\%) on a horizontal plane $17.5~m$ below the free surface under ebb-tide conditions. In (c), two Memorial Bridge piers are visible, and the iso-surfaces of TI$= 12\%$ are shown in red. Flow is from left to right.}
  \label{Fig:3}
  \label{fig:3}
\end{figure}

\noindent We first present the LES-computed wake flow field results of the tidal turbine layout 3 that is illustrated in \prettyref{Fig:2}(d). This LES computed flow field serves as the test case for the ML framework in \prettyref{sec:rad}. \prettyref{fig:3}a depicts the LES-computed turbulence kinetic energy ($k$) normalized by reference velocity ($k/U_{\infty}^2$), together with streamlines of the time-averaged velocity field at the water surface under the ebb-tide conditions. The streamlines reveal a pronounced flow separation along the reach, with a single large recirculation zone dominating all other flow features. As seen in the $k$ distribution, the highest values of $k$ occur within the recirculation zone and in the wakes of the bridge piers, where elevated $k$ persists along the streamlines farther downstream. The flow dynamics within the study area are therefore governed primarily by two features: Badger's Island and the Memorial Bridge piers. Note that the inflow conditions were obtained by dynamically downscaling a previous physics-based model of the Piscataqua River, and for more details on that the reader is referred to \citet{AKSEN2026105302}. One consequence of the inflow conditions is the elevated $k$ along the left bank, where $k/U_{\infty}^2$ peaks at 0.07. We should note that the left bank is designated with respect to the ebb tide condition, and Badger's Island is located on the left bank. Consistent with \citet{AKSEN2026105302}, the elevated turbulence kinetic energy is induced by river bathymetry and, more importantly, the local channel curvature or planform geometry of the river upstream of the island. Downstream of the island and before the flow reaches the Memorial Bridge, the interaction between the incoming ebb flow and the island-induced curvature leads to the formation of a strong recirculation zone with a maximum width of nearly $145~m$, i.e.,  $\sim60\%$ of the local channel width of $\sim240~m$, and a length of $\sim250~m$ (see \prettyref{Fig:3}b).

Next, we examine the footprint of MHK turbines on the LES-computed flow field results of \prettyref{Fig:3}c, since the iso-surface of turbulence intensity field in \prettyref{Fig:3}(c) is largely insensitive to the layout. As seen, this figure presents the LES-computed turbulence intensity that is defined here as $TI=\sqrt{2k/3}/U_{\infty}\times100\%$, on a horizontal plane 17.5 m below the free surface under ebb-tide conditions. Each turbine has a hub height of $5~m$ above the local bed. Because the bathymetry varies across the array, the hub depth below the free surface differs from turbine to turbine, and the plane of $Z=17.5~m$ coincides with hub height only where the local depth is $22.5~m$. \prettyref{tab:locationturbines} lists, for each turbine, the local depth and the resulting offset between the plane and the hub, expressed in meters. We plot in red the volumetric iso-surfaces bounding the regions where TI exceeds $12\%$. If seen from top view, the iso-surfaces in red would coincide with the high turbulence kinetic energy regions identified at the surface of \prettyref{Fig:3}a, namely the shear layer shed from Badger's Island and the wakes of the Memorial Bridge piers. This correspondence suggests that both vortical flow structures, being anchored by the bathymetry and the pier geometry, remain vertically coherent over most of the water depth. The 12\% iso-surface off the island is consistent with the recirculation zone retaining this highly unsteady flow rather than flushing it downstream. The pier iso-surfaces, by contrast, are narrow and extend vertically from the river bed to the free surface, consistent with the depth-spanning blockage imposed by the piers. In the color map of turbulence intensity of \prettyref{Fig:3}(c) , the MHK turbine signature is visible as TI  increases from the ambient values of $3\%$--$5\%$ to more than $10\%$ in the turbine wake region. This high TI footprint, however, is weak relative to that of the bathymetry-driven structures. More importantly, the turbine wakes and the momentum deficit region associated with the wakes do not convect straight downstream; instead, they are deflected by the recirculating flow and bend toward the left bank.

\begin{table}[H]
\centering
\caption{Vertical offset ($\Delta Z$) between the hub height of each turbine (T1–T10) and the horizontal reference surface shown in \prettyref{Fig:3}b. Negative values indicate that the hub lies below the horizontal reference surface (i.e., at a greater water depth), positive values indicate that it lies above the horizontal reference surface, while zero indicates that the reference surface passes through the turbine hub. For all turbines, the reference surface lies within the rotor swept area.}
\label{tab:locationturbines}
\begin{tabular}{ccccccccccc}
\hline
\textbf{Variable} &
\textbf{T1} &
\textbf{T2} &
\textbf{T3} &
\textbf{T4} &
\textbf{T5} &
\textbf{T6} &
\textbf{T7} &
\textbf{T8} &
\textbf{T9} &
\textbf{T10} \\
\hline
$\Delta Z$ (m) & $-1$ & $-1.5$ & $-0.25$ & $-0.5$ & $0.25$ & $0.5$ & $0$ & $0$ & $-1.25$ & $-1.25$\\
\hline
\end{tabular}
\end{table}

\begin{figure} [H]
  \includegraphics[width=1\textwidth]{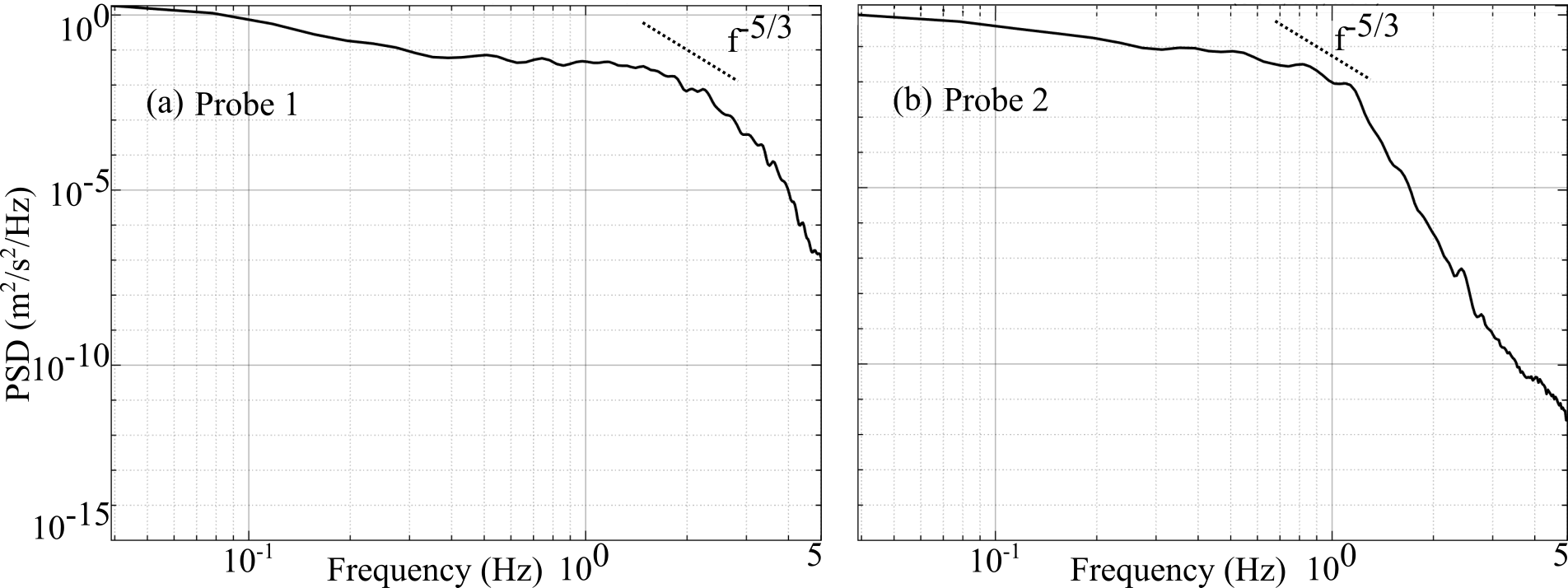} 
  \caption{Power spectral density of the streamwise velocity fluctuations for layout 3 under ebb-tide conditions at (a) probe '$\mathrm{P_1}$' and (b) probe '$\mathrm{P_2}$', as seen in \prettyref{Fig:3}a. Spectra were computed from 6000 samples recorded at approximately $43.2~Hz$ over the period $140~s$ of the statistically converged flow field.}
  \label{Fig:4}
  \label{fig:4}
\end{figure}

In \prettyref{Fig:4} we examine the power spectral density (PSD) of the streamwise velocity fluctuations of the LES-computed flow field for the layout 3 scenario under ebb-tide conditions at probes '$\mathrm{P_1}$' and '$\mathrm{P_2}$', as seen in \prettyref{Fig:3}a. Point '$\mathrm{P_1}$' is located at the edge of the recirculation zone, while point '$\mathrm{P_2}$' is situated between the bridge piers, so that the spectral content of the high turbulence kinetic energy regions can be compared. As seen, both spectra mark a flat, energy-containing regions, which is also known as the production range, at low frequencies that carries most of the resolved variance. The production range is followed by a knee that marks the inertial subrange with a slope of $-5/3$ and then a steep roll-off that represents the dissipation range. 

The two probes differ principally in the location of the knee. More specifically, at point '$\mathrm{P_1}$' the plateau persists to approximately $2~ Hz$, whereas at point '$\mathrm{P_2}$' the roll-off begins near $1~Hz$. Similar results for peak spectral frequencies at river flows are reported in \citet{PSDspectra}. The fluctuations at point '$\mathrm{P_2}$' are therefore dominated by larger and slower flow structures, while point '$\mathrm{P_1}$' retains comparatively more energy at smaller scales, consistent with its position at the edge of the recirculation zone, where the shear layer is thinner and more strongly strained.  The (-5/3) slope marks the inertial subrange, where energy is transferred from larger eddies to smaller ones.

In the dissipation range, where the frequencies are greater than the $-5/3$ slope, turbulent eddies that are smaller than the grid resolution are not directly resolved by LES, but their effect is computed by the subgrid-scale model. The cutoff frequency, below which LES cannot capture the turbulent eddies, is found using Taylor's hypothesis. According to this hypothesis, the cutoff frequency is expressed as $f_{N}=U_{\infty} / (2\Delta s)$, where $\Delta s$ is the spatial resolution. As seen in \prettyref{Fig:4}, at both points, the cutoff frequency is about 4 Hz, meaning that the eddies with frequencies greater than 4 Hz are unresolved.

\subsection{Proper orthogonal decomposition results}
\label{sec:podresults}
\noindent We employed the POD principles to decompose and investigate the contribution of various turbulence modes of the LES-computed flow fields in a study area within the Portsmouth Harbor, NH, as depicted in \prettyref{Fig:2}. Examining the POD results allowed us to determine the contributions of different modes to the energy content of the turbulent flow and  POD's temporal coefficient. We should note that the goal is to detect and isolate the low-order modes, i.e., the coherent flow structures that are inherent to the natural bathymetry and planform geometry of the tidal site. The low-order mode flow structures consist of low-frequency (slowly evolving) large-scale eddies. Once isolated, these energetic low-order modes of turbulence will be utilized to train the proposed machine learning framework, as discussed in \prettyref{sec:pod_unet}.

\begin{figure} [H]
  \includegraphics[width=1\textwidth]{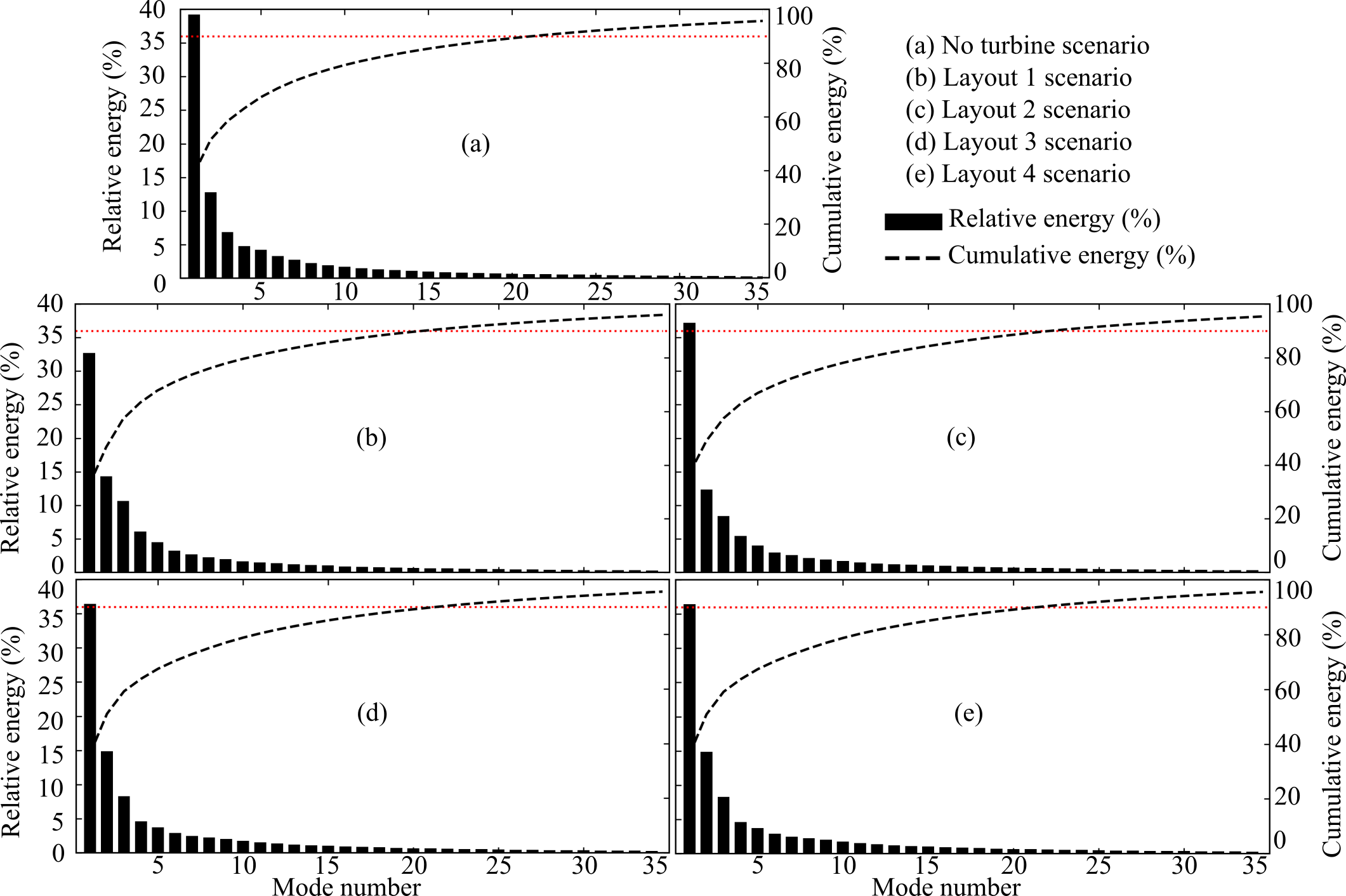} 
  \caption{Individual and cumulative contributions of the first 35 POD modes to the flow energy in (a) the baseline case without turbine array; (b) turbine layout 1; (c) turbine layout 2; (d) turbine layout 3; and (e) turbine layout 4 scenarios. The black dashed line shows the cumulative energy of modes, while the red dotted line shows 90\% of the total turbulence energy of each layout.}
  \label{Fig:5}
  \label{fig:5}
\end{figure}

\prettyref{fig:5} compares the POD energy distributions of the flow in the four turbine array layouts with that of the no-turbine scenario. As seen, each bar shows the energy content of that mode; the dashed lines show the cumulative energy, and the red dotted lines mark the $90\%$ threshold of cumulative energy. As shown, the contributions of modes to the flow energy decrease with increasing mode order. In the baseline case, which lacks turbine arrays (as seen in \prettyref{fig:5}a), the first mode contains approximately $39\%$ of the energy, and it's roughly three times that of the second mode at approximately $13\%$. Also, it shows that nearly $90\%$ of the energy of the baseline case is captured with 21 modes. Comparing the baseline case with the other four cases with turbine arrays, it can be readily seen that the presence of turbine arrays does not substantially alter the overall convergence of the spectra, with 20–23 modes required to reach $90\%$ of the energy. Deploying the turbine arrays, however, seem to reduce the share of energy in the first mode to about $32.5\%$ and $37\%$ across the four trial layouts. It is also evident that the turbine array deployment slightly increases the share of energy content in modes 2 and 3. As such, the combined energy of the first three POD modes remains nearly constant as nearly $57\%$ to $60\%$ across all cases. Thus, the turbine arrays appear to redistribute energy among the leading large-scale structures rather than augmenting their energy level. These isolated low-order modes of turbulence, i.e., the dominant bathymetry-induced coherent flow structures, are directly used to train the machine learning model more effectively.

\begin{figure} [H]
  \includegraphics[width=1\textwidth]{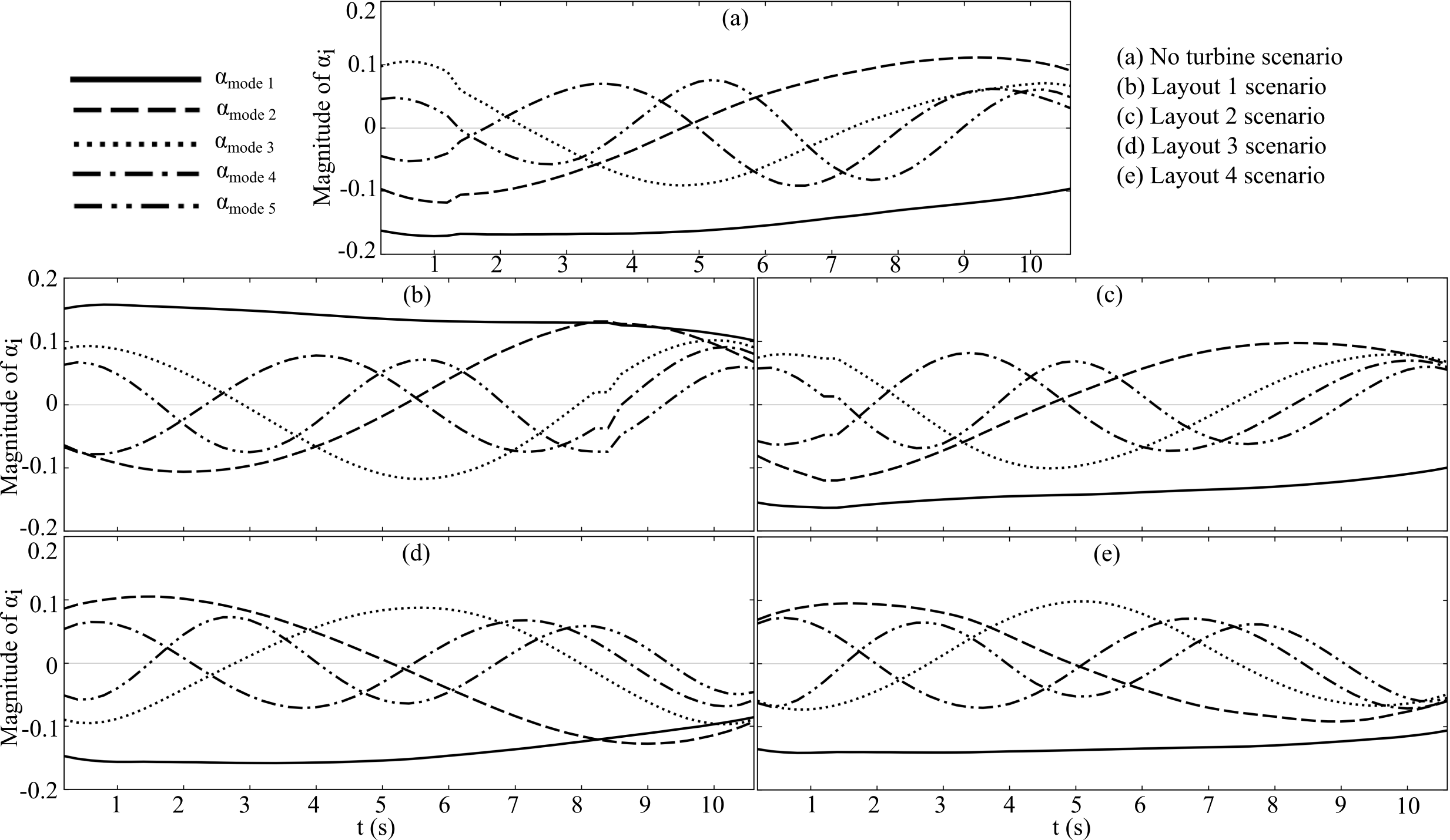} 
  \caption{Time variation of the temporal coefficient of eigenvalue-based models for the first five modes:  baseline case without turbine array, (b) layout 1, (c) layout 2, (d) layout 3, and (e) layout 4. Solid-, dashed-, dotted-, dashed-dotted, and dashed double-dotted lines indicate modes 1, 2, 3, 4, and 5, respectively.}
  \label{Fig:6}
  \label{fig:6}
\end{figure}

To better understand the mode decomposition of the turbulent flow in the study area, we focus on the quantitative analysis of the time variation of the temporal coefficients, $\textit{a}_i(t)$, as described in \prettyref{eq:vel-fluc}. This analysis is performed only for the first five modes of turbulence, which correspond to the large-scale and slow-moving eddies (see \prettyref{Fig:6}). The time history of the modes are shown for a period of $10.5~s$, which is equal to the period of the largest eddy in the waterway ($\mathrm{H_{max}}/U_{\infty}$). As seen, the number of waves observed for each mode increases with mode number, from half a cycle for mode~2 to two cycles for mode~5, since ascending-order modes arrange lower-energy, higher-frequency coherent flow structures in higher modes. Importantly, the range of the absolute values of temporal coefficients of these five modes in the baseline case of layout 1 follows those of other cases closely. We note that the negative and positive temporal coefficients do not indicate reversed flow structures. In other words, the signs returned by the eigenvalue solver carry no physical meaning, as the POD method uses the singular value decomposition transformation, which has an inherent ambiguity in the signs of the eigenvectors \cite{sign}. Thus, once this sign ambiguity is accounted for, the temporal behavior of the leading modes is broadly similar across all five cases, with comparable amplitudes and time scales.
This is consistent with the similar energy distributions in \prettyref{fig:5} and supports that, regardless of turbine layouts the river dynamics can be predominantly determined by the large-scale energetic turbulent eddies.

\begin{figure} [H]
  \includegraphics[width=1\textwidth]{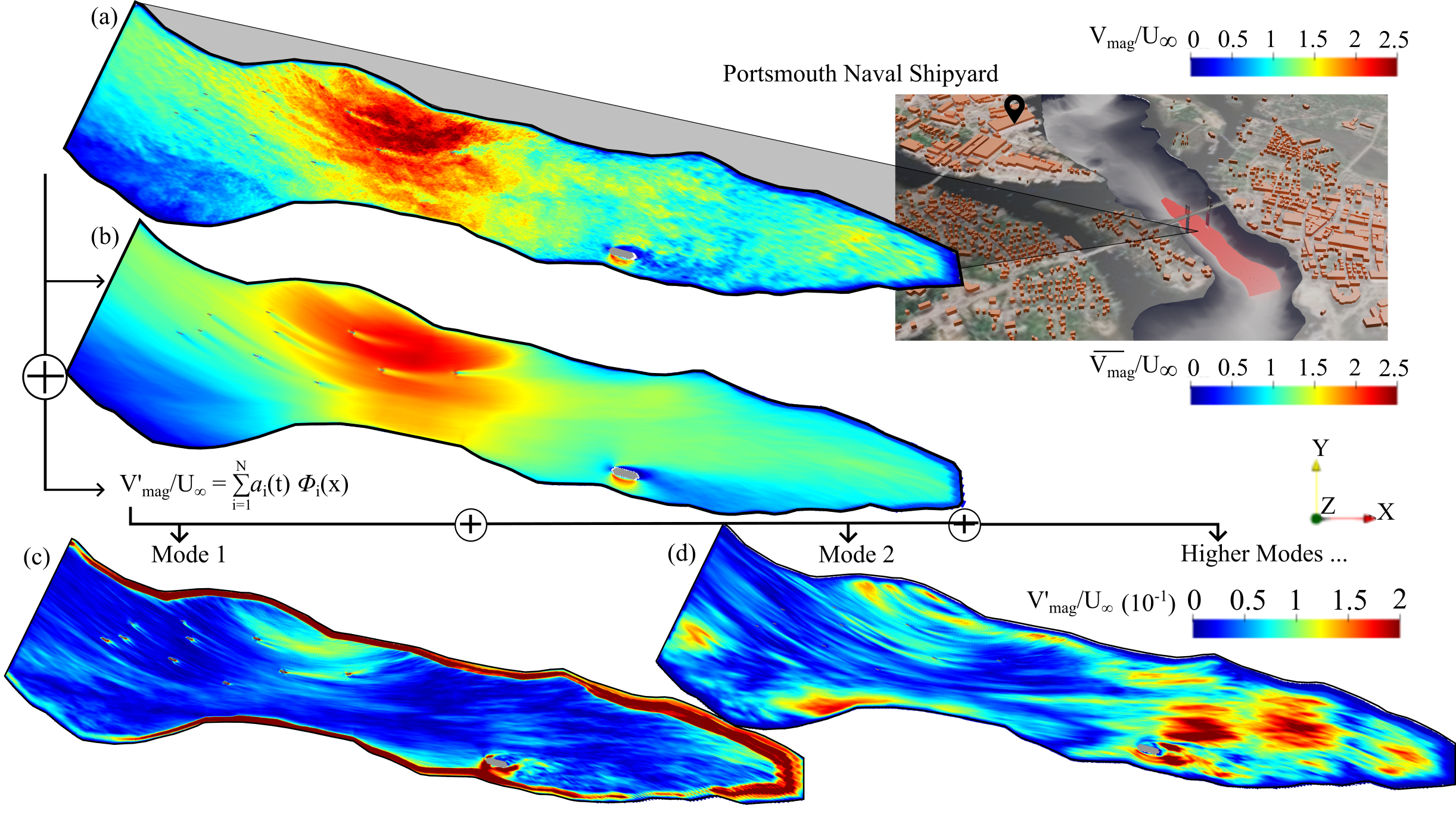} 
  \caption{Colormap of the velocity field and POD modes 1 and 2 over a horizontal plane located $17.5~m$ below the free surface during ebb tide for layout 2.  (a) presents the LES-computed instantaneous velocity magnitude, $\mathrm{V_{mag}/U_\infty}$, at $t = 45~s$. The location of the horizontal plane is marked with the red region of the study area. (b) shows the LES-computed time-averaged velocity magnitude, $\overline{\mathrm{V_{mag}}}/\mathrm{U_\infty}$. The fluctuating component, $u' = \sum_{i=1}^{N} a_i \phi_i$, is reconstructed by the POD modes 1 ($a_1\phi_1$) and mode 2 ($a_2 \phi_2$) in (c) and (d), respectively.}
  \label{Fig:7}
  \label{fig:7}
\end{figure}

To elucidate these modes of POD, in \prettyref{fig:7} we plot the colormap of velocity magnitude captured in modes 1 and 2. The contours of velocity magnitude are shown over a horizontal plane located $17.5~m$ below the free surface for layout 2 under ebb-tide conditions. In this figure, the LES-computed instantaneous velocity field (at $t=45~s$) of panel (a) is broken down into the LES-computed time-averaged (mean) velocity field of panel (b) and the velocity field reconstructed for the first two modes, $a_1 \phi_1$ and $a_2 \phi_2$ of panels (c) and (d), respectively. As seen,  the first mode has clearly captured the wake dynamics of individual turbines and the large-scale mixing layers near the side banks of the tidal river. Consistent with the slow monotonic variation of $a_1$ over the sampling period, mode 1 represents large-scale, low-frequency modulation of the shear layers (\prettyref{fig:6}).

Hence, we argue that the turbine wake dynamics are contained in the first mode and that this has important implications for the training of the proposed machine learning framework. More specifically, our POD analysis revealed that any ML framework built on a truncated POD basis must retain the mode 1 of POD in order to preserve the principal signature of the MHK turbines. In contrast, the second mode consists of elongated, alternating patches distributed throughout the study area. As the second leading mode of turbulence, mode 2 results deem to resemble the turbulence intensity pattern of the flow in \prettyref{Fig:3}c, although with an absence of turbine wake signatures.

\section{POD-filtered 3D U-Net framework}
\label{sec:rad}
\noindent In this section, using the LES results as the benchmark, we compare the performance of two ML frameworks for time-averaged wake flow predictions: a standard 3D U-Net \cite{w18121500, CNN_1, CNN_2, CNN_3, CNN_4} and the proposed POD-filtered U-Net (POD-U-Net), through qualitative and quantitative analysis of the time-averaged velocity field (first-order turbulence statistics). To that end, We first trained the ML models using 18 LES-computed snapshots corresponding to the three turbine layout scenarios (i.e., 6 snapshots from each of the layout 1, 2, and 4). Then, the trained networks were tested using 6 snapshots corresponding to the remaining case, i.e., layout 3 ( \prettyref{Fig:2}b--e). We note that a separate LES was performed for layout 3 to provide a benchmark wake flow to test the trained ML models. This LES-computed benchmark wake flow was not seen by the neural networks during the training process.

Both the standard 3D U-Net and POD-U-Net models have a tri-channel network architecture containing the streamwise ($u$), spanwise ($v$), and vertical ($w$) velocity components. During the training process, the LES-computed time-averaged velocity field (i.e., $\overline{u}, \overline{v}, \overline{w}$) serves as the supervisory target, establishing a many-to-one mapping from its corresponding instantaneous flow fields. As discussed below, both neural networks learned the mapping from instantaneous LES-computed velocity field (input vector) to their corresponding time-averaged flow field (target vector), based on the scheme illustrated in \prettyref{Fig:pod_cnn_framework}d, and feature similar architectures with 96,767 trainable parameters. However, the two neural networks differ fundamentally in the preconditioning of their input data. Namely, the POD-U-Net model projects each input snapshot onto a POD basis and removes a number of selected POD modes of the turbulent flow and subsequently feeds the remaining information into the convolutional layers (as expressed in \prettyref{eq:pod_selective_filter}).

As discussed above, the POD analysis demonstrated that mode 1 must be retained during POD-U-Net training. However, the optimal selection of higher-order modes to filter out remained an open question. To identify the most effective filtering strategy, we systematically evaluated different combinations of filtered POD modes and assessed the resulting POD-U-Net performance against the LES reference data. Specifically, we investigated eight filtering configurations by excluding modes (i) 2–3, (ii) 2–4, (iii) 2–5, (iv) 2–6, (v) 2–7, (vi) 2–8, (vii) 2–10, and (viii) 2–48, while retaining all other modes for network training. The results revealed that filtering modes 2–7 while retaining mode 1 and all modes from 8 onward yielded the best overall POD-U-Net performance. This finding identifies an effective preprocessing strategy for suppressing selected POD modes while preserving the flow structures necessary for accurate turbine wake prediction.

Moreover, our analysis shows that modes 2--7 account for approximately $37\%$ of the resolved turbulent flow structures in the study area. The underlying POD basis was constructed from 53 snapshots sampled at time intervals of $1~s$. Finally, during the final output reconstruction phase, POD-U-Net superimposes the previously subtracted POD modes back onto the network's convolutional prediction.

Furthermore, our analysis indicates that POD modes 2–7 collectively account for approximately $37\%$ of the resolved turbulent flow structures within the study area. The POD basis was constructed using 53 flow-field snapshots sampled at $1~s$ intervals. It is important to note that, during the final reconstruction stage, POD-U-Net reintroduces the previously filtered POD modes into the convolutional predictions, thereby restoring these flow-field components to the reconstructed output. In this section, the velocity magnitude reconstructed from the machine learning frameworks, $\left|\widehat{\left\langle\mathbf{u}^{(\ell)}\right\rangle_t}+\sum_{i=2}^{7}a_i(t)\boldsymbol{\phi}_i(\mathbf{x})\right|$, is compared with the time-averaged LES velocity magnitude, $\sqrt{\overline{u}^2+\overline{v}^2+\overline{w}^2}$. For brevity, both quantities are denoted $\mathrm{\overline{V_{mag}}}$.

\subsection{Qualitative analysis of prediction performance of the POD-U-Net}
\begin{figure} [H]
  \includegraphics[width=1\textwidth]{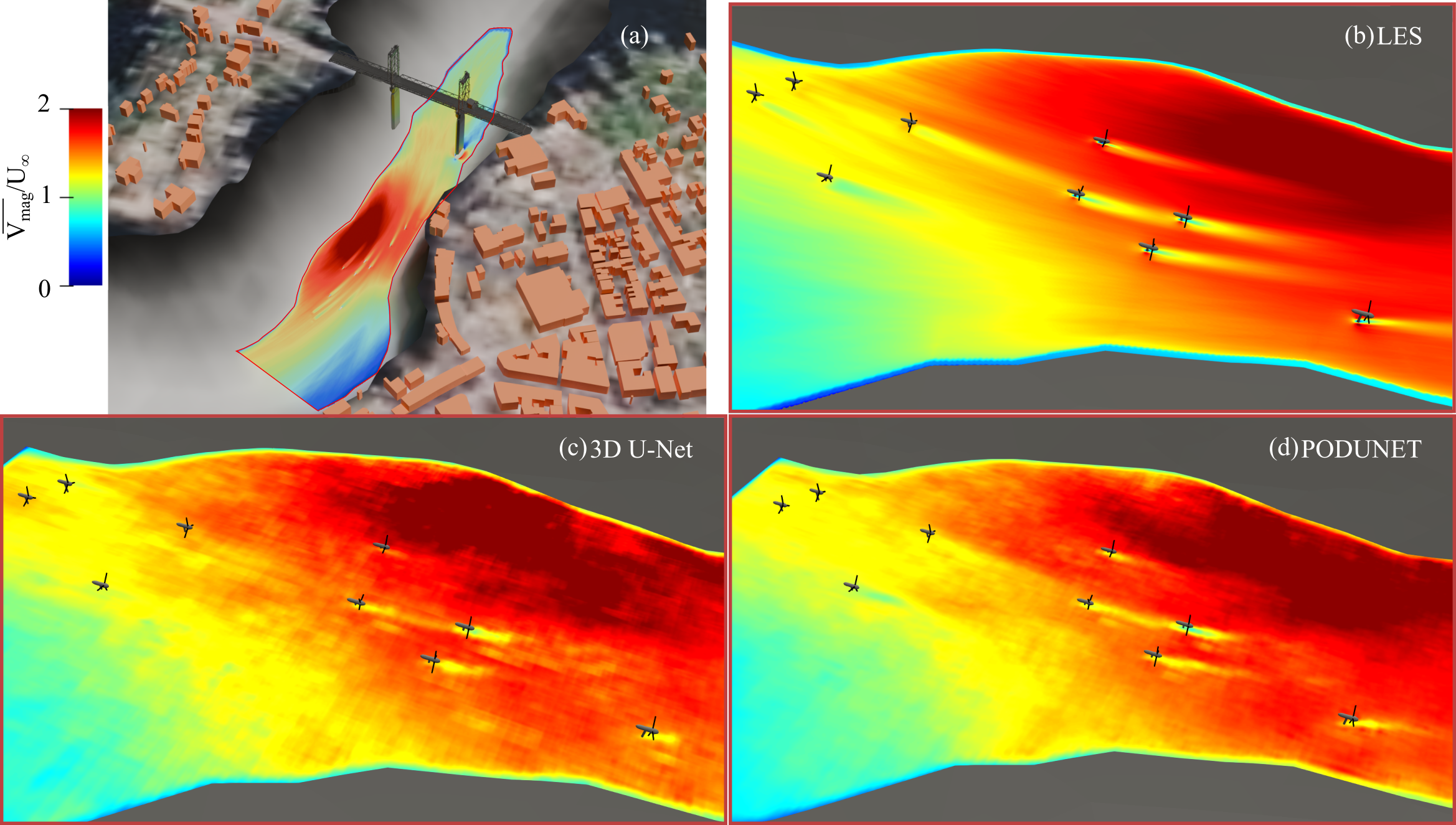} 
  \caption{Color maps of the normalized time-averaged velocity magnitude, $\overline{V_{mag}}/U_{\infty}$, on a horizontal plane located $17.5~m$ below the free surface under ebb-tide conditions for test turbine layout 3. (a) Location of the simulation domain within the study site, with the outlined box indicating the turbine array region shown in greater detail in panels (b)–(d). The normalized velocity fields predicted by (b) LES, (c) the standard U-Net, and (d) POD-U-Net are presented for comparison. In panels (b)–(d), the flow direction is from left to right.}
  \label{Fig:9}
  \label{fig:9}
\end{figure}

\noindent We compare in \prettyref{Fig:9} the LES-computed and machine-learning-predicted time-averaged velocity magnitude fields on a horizontal plane located $17.5~m$ below the free surface under ebb-tide conditions for test turbine layout 3. \prettyref{Fig:9}(a) shows the location of the turbine array within the study site. Consistent with previous investigations of Portsmouth Harbor \cite{AKSEN2026105302, w18121500} and the turbulence intensity distribution presented in \prettyref{Fig:3}, the flow in this region is highly energetic and strongly influenced by the underlying bathymetry, particularly in the wake of Badger's Island. Pronounced velocity deficits immediately downstream of individual turbines indicate energy extraction and the development of turbine wakes. Based on the turbine indexing defined in \prettyref{Fig:2}d, we identify a potential wake–wake interaction between turbines 'T7' and 'T6'. In addition, the wakes generated by turbines 'T8' and 'T9' exhibit pronounced meandering that appears to follow the curvature of the left riverbank, suggesting a strong influence of the local bathymetry on wake evolution. These complex flow features, including wake interactions, wake deflection, and bathymetry-induced meandering, are characteristic of realistic tidal-energy sites. Their coexistence presents a substantial challenge for the standard U-Net and POD-U-Net models, which must distinguish localized turbine-wake signatures from the larger-scale flow structures imposed by the heterogeneous bathymetry to accurately reconstruct the wake flow field.

\prettyref{Fig:9}(c) to (d) plot the predicted velocity field for the standard 3D U-Net and the proposed POD-U-Net, respectively. As seen, the standard 3D U-Net and the proposed POD-U-Net frameworks accurately predict the overall velocity field throughout the study domain. In particular, both surrogate models reliably captured the high-velocity regions near the left bank of the river. Yet, the standard 3D U-Net exhibits incoherent small-scale structures that appear as discrete, patch-like artifacts along the river and most prominently near the right riverbank. Such results with spatial artifacts have also previously been reported in prior studies with standard pixel-wise regression losses in CNNs, particularly when applied to complex fluid flows \cite{oommen2026learning,balasubramanian2026sharper}. We argue that this issue arises due to ($i$) the architecture of these models, which does not penalize continuity of the time-averaged velocity predictions (e.g., see \cite{PINN1, karniadakis2021physics}),  and ($ii$) the limited size of the training dataset \cite{thuerey2020deep, bentivoglio2022deep, sun2020surrogate}. As a result, the standard U-Net overpredicts velocity magnitudes in the vicinity of turbine 'T2'. This localized overprediction is particularly critical for array design and layout optimization, as hydrokinetic power output scales cubically with the incident velocity ($P \propto u^3$). Furthermore, \prettyref{Fig:9}c shows that the standard U-Net fails to reliably reconstruct turbine wake deficits. Namely, the near- and far-field predictions of the standard U-Net wake predictions resulted in overly wide and unnaturally short turbine wakes, which implies underperformance of the model. 

We argue that the energetic tidal flow in real-life ecosystems consists of a wide range of eddy frequencies, which leads to inherently noisy data. The fluctuations of parameters in instantaneous flow fields are governed by large-scale turbulent structures rather than the subtle, localized turbine deficits. These localized inaccuracies underscore the difficulty of the standard regression networks when distinguishing wake interactions from large-scale flow structures. Consequently, the energetic fluctuating content consists of high-variance noise.  As a result, the standard U-Net would effectively suppress these energetic signals to recover accurate mean flow features within the river and way from the tidal turbines. Thus, resolving turbine wakes under these conditions is particularly challenging for CNNs because wake deficits represent subtle, low-energy perturbations embedded within a high-energy background flow.

In contrast, POD-U-Net truncates the dominant fluctuating modes before the input enters the network, effectively filtering out high-variance background turbulence that the network would otherwise suppress internally. It is worth noting that the truncation of specific POD modes was intentionally designed to isolate and emphasize turbine-induced flow structures. With this scope, we decided to retain the modes containing the turbine-induced signal relative to background flow and thereby allowing POD-U-Net to learn to capture turbine wake dynamics. For that, as seen in \prettyref{Fig:9}, POD-U-Net significantly improves the prediction of the momentum deficit throughout the tidal site. While smoothing the input data simplifies the mapping task for the convolutional layers, the post-reconstruction and superposition of the subtracted POD modes restores the fine-scale spatial details as well. Consequently, the POD-U-Net achieves a reasonable mapping compared to the standard U-Net, as the modal superposition process in POD-U-Net prevents the convolution operators from over-smoothing localized spatial gradients. While the predicted near-field wake flow structures of LES and POD-U-Net align closely, minor misalignments remain in the far-field momentum deficit recovery.

\begin{figure} [H]
  \includegraphics[width=1\textwidth]{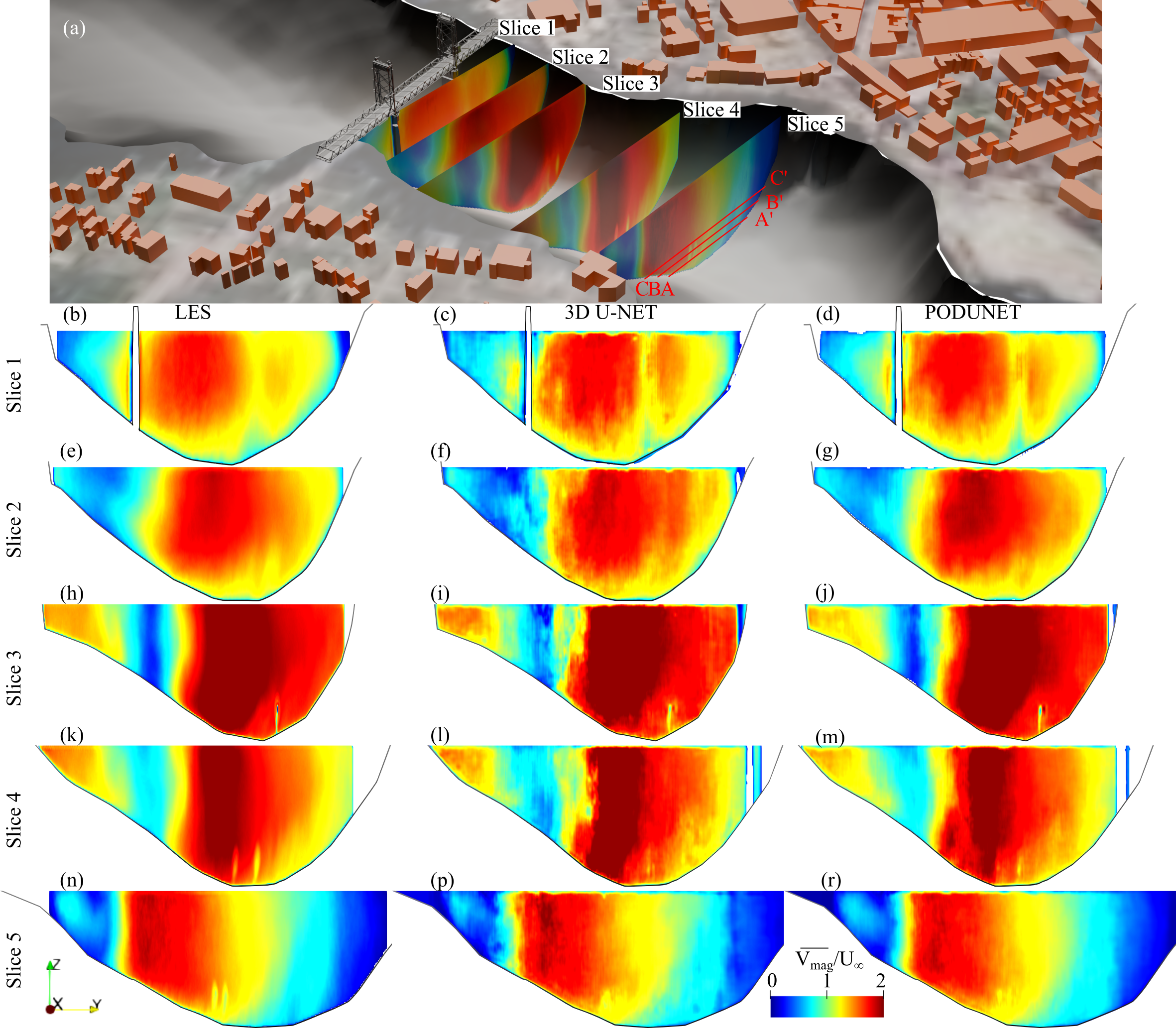} 
  \caption{Color maps of the predicted normalized time-averaged velocity magnitude $\overline{V_{mag}}/U_{\infty}$, over five cross-planes under ebb-tide conditions for the test turbine layout 3. (a) marks the location of cross-planes 1–5 within the study site, where flow is from right to left. (b) to (r) show the velocity fields over the five cross-planes. The rows correspond to sections 1 (top) to 5 (bottom), while the first, second, and third columns correspond to the LES, standard U-Net, and the POD-filtered U-Net predictions, respectively.}
  \label{Fig:10}
  \label{fig:10}
\end{figure}

To further examine these prediction results, we plot in \prettyref{fig:10} the color map of time-averaged velocity magnitude on five cross-planes along the river within the study area. As seen, going downstream from the bridge, there is a redistribution of momentum in the river flow, owing to the infrastructure, planform geometry, and bathymetry of the river in this region. More specifically, the high-velocity core near the bridge shifts from the left bank towards the right bank of the river at the most downstream cross-plane. The overall velocity distribution shows a reasonable prediction for the main river flow for both ML models, compared with the LES. However, the wake dynamics predictions of the two ML models can be compared in the cross-planes of Sections 3 to 5, where there are several MHK turbines installed. As seen, the POD-U-Net has successfully recognized the signature of the turbines relatively well in all cross-planes, while the standard U-Net has barely predicted these turbine wakes. 

\subsection{Quantitative analysis of prediction performance of the POD-U-Net}
\begin{figure} [H]
  \includegraphics[width=1\textwidth]{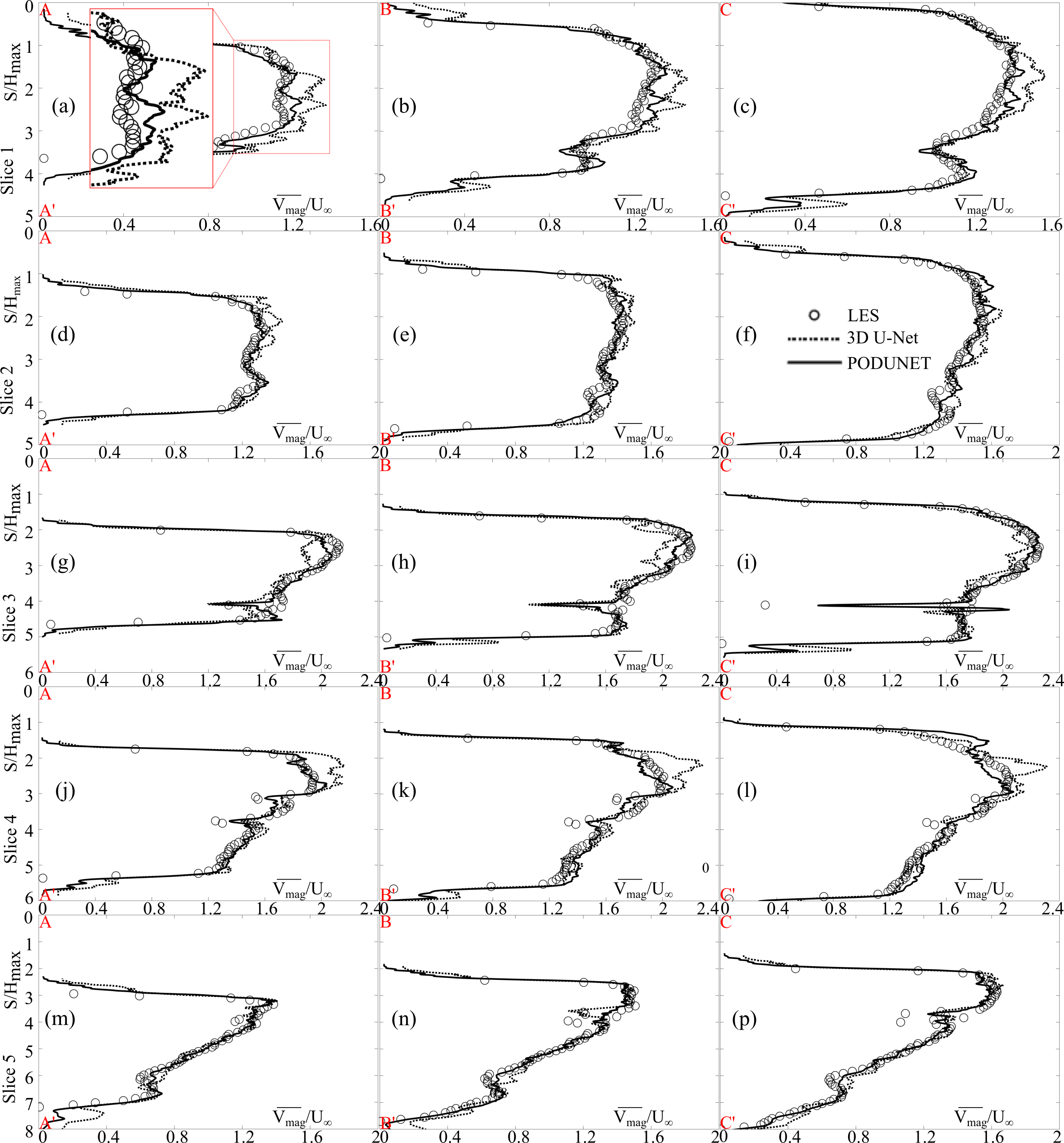} 
  \caption{Transverse profiles of the normalized time-averaged velocity magnitude $\overline{V_{mag}}/U_{\infty}$ along three transect lines for all cross-planes and at water depths of $18.75~m$ (A--A'), $17.50~m$ (B--B'), and $16.25~m$ (C--C'), under ebb-tide conditions for the test turbine layout 3. Rows 1 to 5 correspond to Slices 1 to 5 and columns (from left to right) correspond to transects A--A', 
  B--B, and C--C. Each panel overlays the LES (circles), the standard U-Net prediction (dotted line, and the POD-U-Net (bold line) predictions, as described \prettyref{Fig:10}a. $H_{\mathrm{max}}$ denotes the maximum water depth, (=$25~m$).}  
  \label{Fig:11}
  \label{fig:11}
\end{figure}

\noindent A quantitative analysis was carried out to evaluate the performance of the two ML models against the LES results. To do so, we plot the profiles of veclocity magnitdes across the river and along the horizontal lines A-A', B-B', and C-C' over the five cross-planes, as defined in \prettyref{Fig:10}(a). The tranverse lines A-A', B-B', and C-C' are located in water depths of $18.75~m$, $17.50~m$, and $16.25~m$, respectively. As seen in \prettyref{Fig:11}, compared to the LES results, both ML models predicted the velocity magnitude distributions reasonable at almost all profiles. However, it can be clearly seen that the U-Net model overpridicts the velocity profiles of cross-planes 1 and 4, while missing the momentum deficits at cross-planes 3 to 5.  

To quantitatively assess the performance of the two ML models, we compared their predictions against the LES results by examining velocity-magnitude profiles across the river along the transverse lines A–A', B–B', and C–C' on five cross-planes, as defined in \prettyref{Fig:10}(a). These lines are located at depths of 18.75 m, 17.50 m, and 16.25 m, respectively. As shown in \prettyref{Fig:11}, both ML models reproduce the velocity-magnitude distributions reasonably well across most of the examined profiles. However, the standard U-Net exhibits notable discrepancies, overpredicting velocity magnitudes on cross-planes 1 and 4 while failing to capture the momentum deficits on cross-planes 3 to 5. These results highlight the limitations of the standard U-Net, compared against the POD-U-Net, in accurately reproducing turbine-induced wake features under the complex hydrodynamic conditions of the study site.

\begin{table}[H]
\centering
\caption{Error metrics of the ML models for the velocity field: mean squared error (MSE), root mean squared error (RMSE), and mean absolute error (MAE). The first three columns show the results for 3D U-Net and the last three columns for POD-U-Net.}
\label{tab:errormetrics}
\begin{tabular}{ccccccc}
\hline
 &
 &
\textbf{3D U-Net} &
 & 
 &
\textbf{POD-U-Net} &
 \\
\hline
\textbf{Metrics} &
\textbf{$\widehat{\left\langle\mathbf{u}^{(\ell)}\right\rangle_t}$} &
\textbf{$\widehat{\left\langle\mathbf{v}^{(\ell)}\right\rangle_t}$} &
\textbf{$\widehat{\left\langle\mathbf{w}^{(\ell)}\right\rangle_t}$} & 
\textbf{$\widehat{\left\langle\mathbf{u}^{(\ell)}\right\rangle_t}$} &
\textbf{$\widehat{\left\langle\mathbf{v}^{(\ell)}\right\rangle_t}$} &
\textbf{$\widehat{\left\langle\mathbf{w}^{(\ell)}\right\rangle_t}$} \\
\hline
$MSE$ & 0.0212 & 0.0072 & 0.0004 & 0.0044 & 0.0019 & 0.0002 \\
$RMSE$ & 0.1456 & 0.0851 & 0.0187 & 0.0662 & 0.0438 & 0.0157 \\
$MAE$ & 0.1189 & 0.0470 & 0.0115  & 0.0410 & 0.0282 & 0.0095 \\
$\mathrm{R^2}$ & 0.9702 & 0.9404 & 0.4867 & 0.9938 & 0.9842 & 0.6389  \\
\hline
\end{tabular}
\end{table}

Moreover, \prettyref{tab:errormetrics} presents the error metrics of the 3D U-Net and POD-U-Net architectures for the time-averaged velocity components. Performance was evaluated using MSE, RMSE, MAE, and the coefficient of determination ($R^2$). Overall, POD-U-Net demonstrated superior predictive performance compared to the standard 3D U-Net across all metrics. Specifically, for the primary velocity component ($\overline{u}$), POD-U-Net achieved one-order of magnitude  reduction in error relative to the 3D U-Net predictions, while elevating $R^2$ from $0.9702$ to $0.9938$. Likewise, similar enhancement can be observed for other velocity components.

Furthermore, \prettyref{tab:errormetrics} summarizes the performance of the 3D U-Net and POD-U-Net architectures in predicting the time-averaged velocity components, using four standard evaluation metrics: MSE, RMSE, MAE, and the coefficient of determination ($\mathrm{R^2}$). Overall, POD-U-Net consistently outperforms the standard 3D U-Net across all metrics. For the primary velocity component ($\overline{u}$), POD-U-Net achieves an approximately one-order-of-magnitude reduction in prediction error compared with the standard 3D U-Net, while increasing $\mathrm{R^2}$ from 0.9702 to 0.9938. Comparable improvements are observed for the other two velocity components, further demonstrating the effectiveness of the POD-based filtering strategy in enhancing the predictive accuracy of the neural network.

\begin{figure} [H]
  \includegraphics[width=1\textwidth]{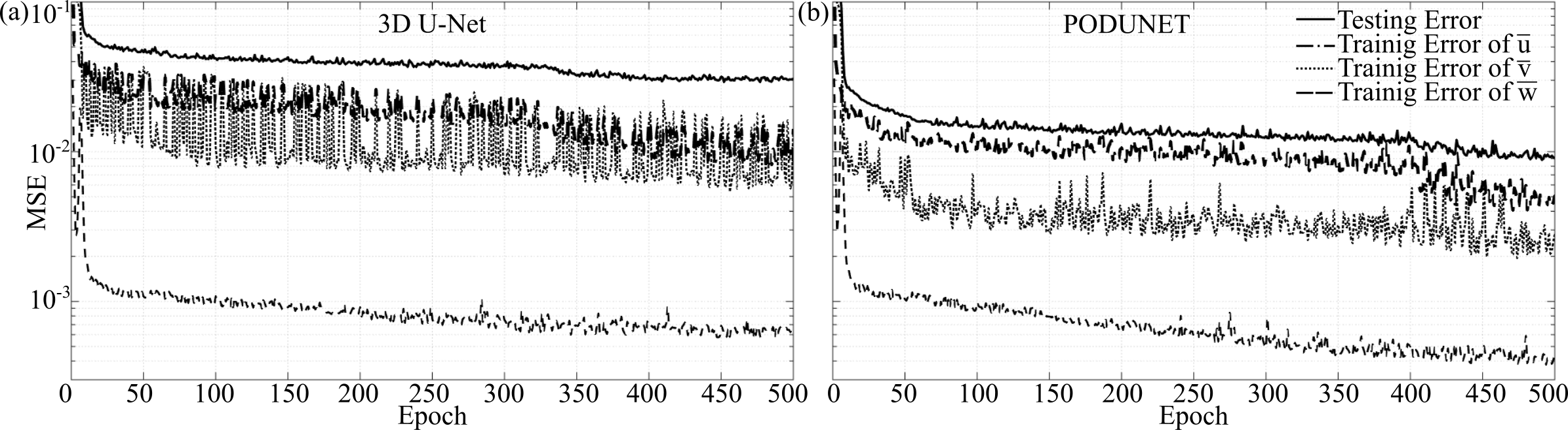} 
  \caption{Training and testing MSE histories over training epochs for a) Standard 3D U-Net framework and b) POD-filtered 3D U-Net framework, plotted on a logarithmic scale. The solid, dashed-dotted, dotted, dashed lines refer to testing error, and errors corresponding to u, v, and w, respectively.}
  \label{Fig:12}
  \label{fig:12}
\end{figure}

To further evaluate the learning performance of the two neural networks, \prettyref{fig:12} compares their training and testing mean squared error (MSE) histories. The testing error is reported for the velocity magnitude, whereas the training errors are presented separately for the three velocity components. As shown in the figure, both models exhibit similar overall convergence trends; however, POD-U-Net consistently outperforms the standard U-Net across all evaluated metrics. For both models, the error decreases sharply during the first 10–20 epochs, followed by a more gradual improvement. Neither model exhibits an increase in testing error during the later training stages, suggesting that neither experiences evident late-stage divergence. Nevertheless, the standard U-Net converges relatively early, with its final testing error leveling off at approximately $3.5 \times 10^{-2}$. In contrast, POD-U-Net continues to improve throughout training, achieving a final testing error of approximately $1.0 \times 10^{-2}$. This represents a reduction of approximately $71\%$ in the final testing error relative to the standard U-Net. A similar improvement in convergence and error reduction is observed for each of the three velocity components, further demonstrating the enhanced learning performance of POD-U-Net.

Lastly, as seen in \prettyref{fig:12}, POD-U-Net exhibits a more stable training process than the standard U-Net, as evidenced by the substantially smaller epoch-to-epoch fluctuations in the training errors of all three velocity components. The difference between the training and testing errors is also noticeably smaller for POD-U-Net, suggesting improved generalization performance. These improvements can be attributed to the POD-based filtering strategy, which suppresses selected flow structures that are potentially difficult for the neural network to learn. By removing these components from the learning process, POD filtering provides a smoother and more learnable representation of the target flow field, enabling the proposed POD-filtered U-Net model to achieve more stable training and improved predictive accuracy.

\section{Conclusion}

\label{sec:conc}

\noindent We developed a POD-filtered 3D U-Net (POD-U-Net) surrogate to predict the time-averaged flow field of tidal turbine arrays in Portsmouth Harbor, New Hampshire, under ebb-tide conditions. The Piscataqua River served as the site of interest, building on our previous feasibility study \cite{AKSEN2026105302}, which identified the region between Badger's Island and the Portsmouth Naval Shipyard as the most favorable hypothetical region for deploying 5 m rotors within the navigation constraints of the harbor's deep-draft channel. A 1 km-long array-scale flow domain was dynamically downscaled from the $6~km$-long site-scale flow domain through one-way nesting of its inflow boundary conditions. Four layouts, each with ten installed MHK horizontal axis turbines, parametrized by the actuator line method, were modeled with LES to generate the training and testing dataset to train and validate the two ML models. Performing POD of the flow field data, we sought to separate the bathymetry-induced coherent flow structures from the turbine-induced flow features before training, so that the network could concentrate its learning on the latter.

The LES results of the study area demonstrate that the flow in this region is primarily influenced by the planform geometry and bathymetry of the river, as well as the hydraulic structures rather than by the turbines. Namely, the flow separation downstream of Badger's Island generates a recirculation zone approximately 145 m wide and 250 m long, occupying about 60\% of the river width. This recirculation zone determines most dominant structures and river's regime. Although certain turbine wakes raise the turbulence intensity from ambient values of 3–5\% to above 10\%, their overall contributions and, thus, footprint in the resolved LES data is weak relative to the bathymetry-driven structures.

The POD analysis confirms this hierarchy. Introducing turbines reduces the share of the first mode of the turbulence kinetic energy. Yet in all cases the first three modes retain a nearly constant 57–60\%, and 20–23 modes are required to reach 90\%. The temporal coefficients of the leading modes are also similar across all layouts. Hence, the turbines redistribute energy among the leading large-scale structures rather than adding to it, and the dominant coherent structures are intrinsic to the river. Notably, the first mode also carries the shear layers generated by the turbines, whereas the second mode shows no turbine signature. It is therefore necessary to retain the first mode when filtering, rather than truncating the leading modes by energetic rank alone.

Building on this understanding, modes 2–7 of the turbulence were filtered during the training process (and before the convolutional stage) of the POD-U-Net model. The network architecture and loss function of the two ML models remain identical. Both surrogates were trained on three MHK turbine array layouts and tested on an unseen fourth, and both reproduce the overall velocity distribution of the LES. However, the standard 3D U-Net produces patch-like artifacts near the right bank and across flow-separation regions and overestimated the velocity field in the vicinity of some turbines, returning overly thick and short turbine wakes. The POD-U-Net more faithfully recovers the turbine wake deficits, the recirculation zone along the left bank, and the bridge-induced shear layer.



\section{Acknowledgements}
\label{sec:acknowledge}
\noindent This work was supported by grants from the National Science Foundation (grant number 2233986) and the U.S. Department of Energy’s Office of Critical Minerals and Energy Innovation under the Hydropower and Hydrokinetic Office, Award Numbers DE-EE0009450 and DE-EE00011379. The views expressed herein do not necessarily represent the views of the U.S. Department of Energy or the United States Government. The computational resources for the simulations of this study were partially provided by the Institute for Advanced Computational Science at Stony Brook University. 

\section{Data Availability Statement} 
\label{sec:7} 
The code for the numerical model (10.5281/zenodo.10521565), including the simulation results for the flow field (10.5281/zenodo.23190082), and code for POD-U-NET (10.5281/zenodo.23190082) are made available via the Zenodo online repository.

\bibliographystyle{elsarticle-num-names} 
\bibliography{main.bib}
\end{document}